\documentclass[12pt]{article}
\usepackage{epsfig,amssymb}

\newcommand{\bq}{\begin{equation}}
\newcommand{\eq}{\end{equation}}
\newcommand{\bqa}{\begin{eqnarray}}
\newcommand{\eqa}{\end{eqnarray}}
\newcommand{\ben}{\begin{enumerate}}
\newcommand{\een}{\end{enumerate}}
\newcommand{\bc}{\begin{center}}
\newcommand{\ec}{\end{center}}
\newcommand{\bqb}{\begin{eqnarray*}}
\newcommand{\eqb}{\end{eqnarray*}}

\def\gsim{\gtrsim}

\def\pr#1#2#3{ Phys. Rev. ${\bf{#1}}$ (#2) #3}

\def\pl#1#2#3{ Phys. Lett. ${\bf{#1}}$ (#2) #3}
\def\prep#1#2#3{ Phys. Rep. ${\bf{#1}}$ (#2) #3}

\def\np#1#2#3{ Nucl. Phys. ${\bf{#1}}$ (#2) #3}
\def\zp#1#2#3{ Z. f. Phys. ${\bf{#1}}$ (#2) #3}
\def\epj#1#2#3{ Eur. Phys. J. ${\bf{#1}}$ (#2) #3}

\def\cpc#1#2#3{Comput. Phys. Commun. ${\bf{#1}}$ (#2) #3}
\def\nuovo#1#2#3{ Nuovo Cim. ${\bf{#1}}$ (#2) #3}

\def\ie{{\it i.e.\/}}
\def\eg{{\it e.g.\/}}

\def\etal{{\it et.al.\/}}

\global\nulldelimiterspace = 0pt

\def\ii{ \mathrm{i}}
\def\L{ {\cal L }}
\def\H{ {\cal H }}

\def\sw{s_W}
\def\cw{c_W}
\def\swd{s^2_W}
\def\cwd{c^2_W}

\def\mwd{m_W^2}
\def\mw{m_W}
\def\mz{m_Z}
\def\mzd{m_Z^2}
\def\mf{m_f}
\def\mh{m_h}

\def\t{\hat t}
\def\s{\hat s}
\def\u{\hat u}

\def\Fstu{\tilde F(\s, \t, \u)}
\def\Est{E_1(\s, \t)}
\def\Esu{E_1(\s, \u)}
\def\Etu{E_2(\t,\u)}

\begin{document}
\pagenumbering{arabic}
\thispagestyle{empty}
\def\thefootnote{\fnsymbol{footnote}}
\setcounter{footnote}{1}

\begin{flushright}
PM/99-32 \\
THES-TP 99/08 \\
hep-ph/9909243 \\
September 1999
 \end{flushright}
\vspace{2cm}
\begin{center}
{\Large\bf The $\gamma \gamma \to Z Z $ process and the search
for virtual SUSY effects at a $\gamma \gamma $ Collider.}\footnote{
Partially supported by the NATO grant CRG 971470 and
by the Greek government grant PENED/95 K.A. 1795.}
 \vspace{1.5cm}  \\
{\large G.J. Gounaris$^a$, J. Layssac$^b$, P.I. Porfyriadis$^a$ and
F.M. Renard$^b$}\\
\vspace{0.7cm}
$^a$Department of Theoretical Physics, Aristotle
University of Thessaloniki,\\
Gr-54006, Thessaloniki, Greece.\\
\vspace{0.2cm}
$^b$Physique
Math\'{e}matique et Th\'{e}orique,
UMR 5825\\
Universit\'{e} Montpellier II,
 F-34095 Montpellier Cedex 5.\\
\vspace{0.2cm}

\vspace*{1cm}

{\bf Abstract}
\end{center}

We study the helicity amplitudes
of the process $\gamma \gamma \to Z Z$ in the
Standard Model  at high energy. These amplitudes
receive contributions from
the  $W$ and charged quark and lepton loops, analogous to
those encountered in the
$\gamma \gamma \to \gamma\gamma, ~\gamma Z$ cases studied before.
But $\gamma \gamma \to ZZ$  also receives contributions from
the Higgs s-channel poles involving the effective
Higgs-$\gamma \gamma$ vertex. At energies $\gsim 300 GeV$,
the  amplitudes in all three processes
are mainly helicity-conserving and almost purely imaginary;
which renders them
a very useful tool in searching for New Physics.
As an example, a SUSY case is studied,
and the signatures due to the virtual effects induced by a
chargino-,  charged slepton- or a lightest stop-loop in
$\gamma \gamma \to ZZ$, are explored. These signatures,
combined with the analogous ones in
$\gamma \gamma \to \gamma \gamma$
and  $\gamma \gamma \to \gamma Z$, should help identifying
the nature of possible New Physics particles. \par

\def\thefootnote{\arabic{footnote}}
\setcounter{footnote}{0}
\clearpage

\section{Introduction}

\hspace{0.7cm}In the previous papers \cite{gggg, GPRgamma, gggZ}
we have presented a thorough study of
the processes $\gamma \gamma \to\gamma \gamma $ and
$\gamma \gamma \to \gamma Z$ in the Standard (SM) and SUSY models.
These processes  do not appear at tree level, and first arise at
1-loop order. In the Standard Model (SM) at energies
above $250~GeV$, their most striking
property is that they are strongly dominated by the
two independent helicity amplitudes $F_{++++}(\s,\t,\u)$ and
$F_{+-+-}(\s,\t,\u)=F_{+--+}(\s,\u,\t)$, which moreover turn out
to be largely imaginary; the effect being more pronounced
at the smaller scattering angles. At such energies all
the other helicity amplitudes
are extremely small.
This remarkable property is due to the fact that the real
Sudakov-type log-squared terms contributed by the various
1-loop diagrams, cancel out for all physical
amplitudes. As a result, the most important remaining contribution
at high energy and fixed scattering angle,
is due to  the single-log,  predominantly
imaginary   terms, contributed by
the $W$-loop diagrams. These terms only affect the  helicity
conserving  amplitudes. All  other amplitudes
receive comparable contributions from both the
$W$ and  fermion loops, and turn out to be very
small in SM.  Since a similar property is naturally
expected also for the process  $\gamma \gamma \to ZZ$
at sufficient energies, we intend here to present its study. \par

The processes $(\gamma \gamma \to \gamma \gamma ~ , ~\gamma Z
~ , ~ ZZ) $,  could be measured at the future $e^+e^-$ Linear
Colliders (LC) \cite{LC}, when operated as a $\gamma \gamma $ Collider
($LC_{\gamma \gamma }$) through  backscattering of laser beams
\cite{LCgg, gamma97}. In such a case the $\gamma \gamma $ c.m.
energy could  be as high as $80\%$ of the initial $e^+e^-$
c.m. energy, while an annual luminosity of
$\bar L_{\gamma \gamma } \simeq  0.2 \bar L_{ee}
\simeq 100 fb^{-1}$
would  be reasonably expected \cite{gamma97}.
Polarized $\gamma \gamma$ beams can also be
obtained using initially polarized electron beams and lasers. \par

The aforementioned  simplicity of the SM amplitudes for
the three processes
($\gamma \gamma \to \gamma \gamma ~,~ \gamma Z ~,~ ZZ$),
may someday render them a very useful  in the
search  for New Physics (NP) \cite{GPRgamma};
particularly for NP characterized by appreciable imaginary
contributions to the helicity conserving amplitudes \cite{GPRgamma}.
Such effects could involve \eg\@ amplitudes
containing CP violating
phases; or even  effects due to the possible  existence of
additional large space-dimensions, inducing  contributions from
 strings of graviton-   or $Z$- or $\gamma $-Kaluza-Klein states
with, maybe,  appreciable width-generated  imaginary
parts \cite{Cheung, Antoniadis}.\par

As an example of such an NP search, we
studied previously the effects induced by the various SUSY particle
loops contributing to $\gamma \gamma \to \gamma \gamma ~ ,
~\gamma Z$ \cite{gggg, GPRgamma, gggZ}.
In these studies we concentrated on the idea that there
is no CP-violating phase in the SUSY parameter
space\footnote{An investigation of the effects of such
phases we intend to present in the future.};
so that energies above the threshold for the SUSY particle
production are needed, for appreciable imaginary
contributions to occur.
Of course, at such energies,  the SUSY particles will be
also directly produced and studied with much higher statistics.
Nevertheless, the experimental study of their virtual contribution
to the three processes $\gamma \gamma \to \gamma \gamma ~ , ~\gamma Z
~ , ~ ZZ $, should  provide independent information,  which
will help  to identify their nature. Particularly because
such virtual  SUSY effects should in general be less
sensitive  to the soft symmetry breaking parameters,
than the direct production ones. \par

As already indicated, in the present paper  we complete our previous
analysis of
$\gamma \gamma \to \gamma \gamma ~ , \gamma Z$, by also
studying   the $\gamma \gamma \to
Z Z $ amplitudes in the standard and SUSY models.
The distinctive feature of this later process
(as opposed to the previous ones),
is that it also receives
contributions from the Higgs s-channel pole diagrams\footnote{note
that in $\gamma \gamma \to \gamma \gamma ~ , \gamma Z$, the Higgs
resonance contribution is only absent at one-loop order, while it
contributes from two loops onwards.}, which
increase the sensitivity to the lightest stop, making it measurable.
Of course $\gamma \gamma \to ZZ$ has also been studied before
in SM \cite{JikiaZZ, ZZloop, Dicus}, but explicit expressions for the
for the $W$-loop contribution to the SM amplitudes have only by
given by \cite{JikiaZZ}. We have reproduced the results
of these authors\footnote{Apart from a minor misprint in
the small $F_{+-+0}$ amplitude, to be mentioned below.}
in Appendix A, choosing a different way of presentation though.\par

More explicitly, the  expressions for the   $W$
\cite{JikiaZZ} and fermion loop \cite{GloverZZ}
contributions to the helicity amplitudes are given in Appendix A.
In addition,  we also give the  1-loop contribution
induced by a single charged scalar particle.
In Appendix B
simple asymptotic expressions for the SM helicity amplitudes are given,
which elucidate their physical properties at high energies.\par

In Sec.2 we discuss
the main properties of the exact  expressions for the
$W$,  fermion or  scalar particle 1-loop contributions.
 This allows us to  study the helicity amplitudes in SM, and to
predict possible contributions due to new fermion or
scalar particle loops. As an example
we present the contributions to these amplitudes
due to a gaugino- or higgsino-like chargino, an L- or R-slepton, or
a  lightest stop-loop. In all  applications we
assume no CP-violating phases in the soft SUSY breaking
parameters, and work in the so called decoupling regime, where
the CP-odd neutral Higgs
particle is taken very heavy; $m_A^0 \gg \mz$.
Since the  asymptotic expressions for the SM helicity amplitudes,
derived in  Appendix B, may be useful for quick calculations;
we  also offer in Sec. 2 a discussion of their region of
validity.   \par

In Sec. 3, we study the corresponding $\gamma \gamma \to
Z Z $ cross sections
for various polarizations of the incoming photons. We identify the
sensitivity of these cross sections to various SUSY effects and we
discuss their observability.
Finally, in Sec. 4, we summarize the results and
give our general conclusions for all three processes
$\gamma \gamma \to \gamma \gamma ~,~ \gamma Z ~,~ ZZ$.

\section{An overall view of the $\gamma \gamma \to Z Z $
amplitudes.}

\hspace{0.7cm}The invariant helicity amplitudes  $F_{\lambda_1 \lambda_2
\lambda_3\lambda_4}(\beta_Z, \t,\u)$ for the process
$\gamma \gamma \to Z Z $, with $\lambda_j$ denoting the helicities
of the incoming and outgoing particles,
are given in Appendix A. As observed  in \cite{GloverZZ, JikiaZZ},
the properties of the helicity polarization vectors suggest
to describe the energy-dependence of these amplitudes in terms
of the dimensionless  variable  $\beta_Z$ related to the usual
$\s$ through $\s=4\mzd/(1-\beta_Z^2)$.  In the $ZZ$-rest
frame,   $\beta_Z$ describes the velocity of each $Z$,
provided it is chosen to be positive.  According to the
discussion in Appendix A, the
constraint (\ref{-beta}), together with  (\ref{Bose2}-\ref{Bose1})
and (\ref{parity}), arising from Bose symmetry and
parity invariance respectively, reduce the number of independent
helicity amplitudes to just eight. As in (\ref{8basic})
of Appendix A, these are taken to be
\bqa
& F_{+++-}(\beta_Z,\t,\u)~ , ~ F_{++++}(\beta_Z,\t,\u), &
  F_{+-++}(\beta_Z,\t,\u)~, ~ F_{+-00}(\beta_Z,\t,\u)~, \nonumber \\
&F_{++00}(\beta_Z,\t,\u)~ ,~  F_{+++0}(\beta_Z, \t,\u), &
F_{+-+0}(\beta_Z,\t,\u)~ , ~  F_{+-+-}(\beta_Z,\t,\u)~ .
\label{8basictext}
\eqa
As explained in Appendix A, the relations
(\ref{-beta++--}, \ref{-beta++-0}) implied from
(\ref{-beta}), determine through the  $(\beta_Z \to  -\beta_Z)$
substitution,   the two helicity amplitudes
\bqa
F_{++--}(\beta_Z, \t,\u) & = & F_{++++}(-\beta_Z, \t, \u)
\  \ , \label{-beta++--text} \\
F_{++-0}(\beta_Z, \t,\u) & = & F_{+++0}(-\beta_Z, \t, \u)
\  \ , \label{-beta++-0text}
\eqa
while all the rest are  obtained
from the aforementioned ten, through helicity changes or
$(\t \leftrightarrow \u)$ interchanges.\par

In Appendix A, we  reproduce the  $W$
and charged fermion contributions of
\cite{JikiaZZ, GloverZZ} to  the eight basic amplitudes in
(\ref{8basictext}); while in (\ref{FSamp}, \ref{gSZ})) we also
give  the contributions  due to a loop realized by scalar
particle  carrying a definite weak isospin and charge.
All results are given in terms of the standard 1-loop functions
$B_0$, $C_0$ and $D_0$, first introduced in \cite{Passarino}.\par

Explicit asymptotic expressions for these functions,
as well as for the corresponding $W$, fermion  and scalar loop
contributions to the   helicity amplitudes,
are given in Appendix B. On the basis of them we
conclude that in $\gamma \gamma \to ZZ$, (as well as in the process
$\gamma \gamma \to \gamma \gamma ~,~ \gamma Z$ studied before),
the Sudakov-type real log-squared terms always cancel out
at high energies and fixed scattering angle.
The dominant contributions then arise from logarithmically increasing
imaginary terms generated by the $W$ loop.
It turns out that such terms exist only for  the two
helicity conserving amplitudes  $F_{++++}(\beta_Z,\t,\u)$ and
$F_{+-+-}(\beta_Z,\t,\u)=F_{+--+}(\beta_Z,\u,\t)$;
which are therefore the most important ones at high energies.
These dominant amplitudes are largely imaginary and
increase with energy, while all the rest tend asymptotically
to  quite negligible constants. \par

\begin{figure}[p]
\vspace*{-4cm}
\[
\epsfig{file=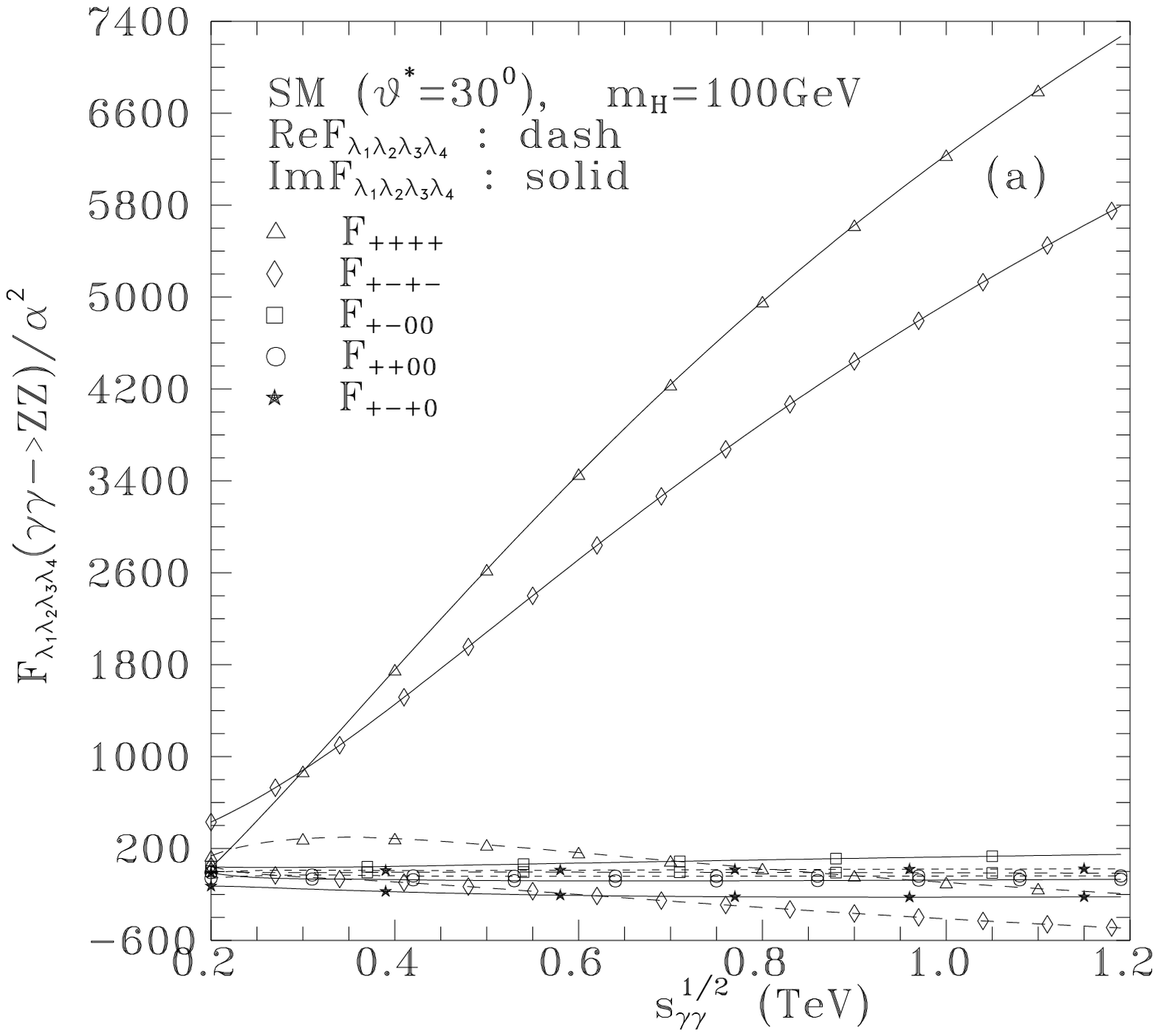,height=7.5cm}\hspace{0.5cm}
\epsfig{file=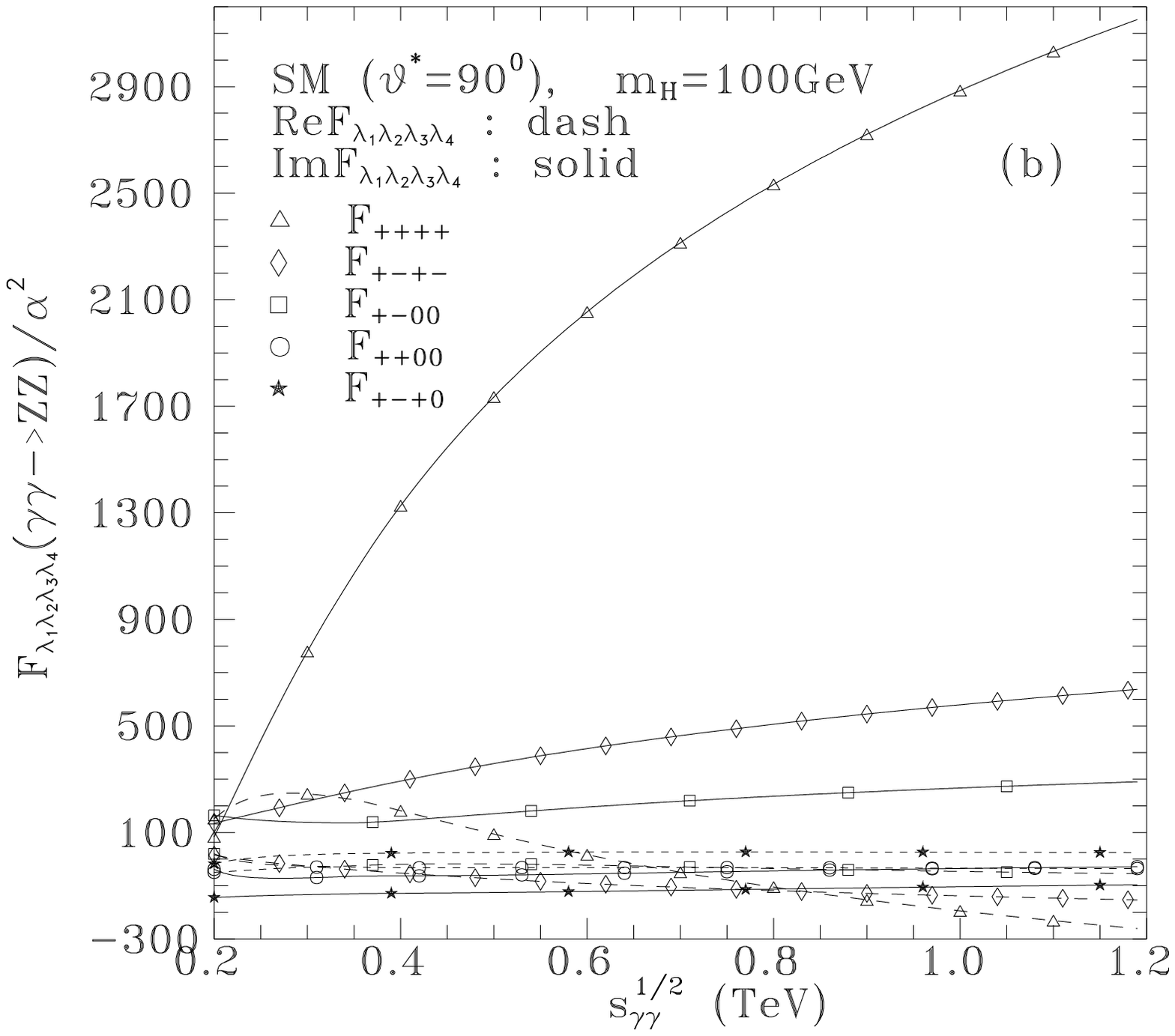,height=7.5cm}
\]
\vspace*{0.5cm}
\caption[1]{SM contribution to the dominant
$\gamma \gamma \to Z Z$ helicity amplitudes at
$\vartheta^*=30^0$ and $\vartheta^*=90^0$. All other amplitudes
are predicted to be smaller or about equal to $F_{+-+0}$.}
\label{SM-amp}
\end{figure}

These results can  be seen in Fig.\ref{SM-amp}a,b,
where the largest among the ten amplitudes in (\ref{8basictext},
\ref{-beta++--text},  \ref{-beta++-0text}) are shown,
using the exact 1-loop functions, at the c.m. scattering angles
$\vartheta^*=30^0$ and $\vartheta^*=90^0$.  It is shown in these figures
that indeed  at sufficient energies, the real parts of
$F_{\pm\pm\pm\pm}(\beta_Z,\t,\u)$ and
$F_{\pm\mp\pm\mp}(\beta_Z,\t,\u) = F_{\pm\mp\mp\pm}(\beta_Z,\u, \t)$
are always much smaller than the corresponding imaginary parts.
The effect becomes less pronounced though, as the
scattering angle increases.\par

We have also checked that for $\sqrt{\s}\gsim 300~ GeV$, the $W$-loop
contribution completely dominates the large imaginary parts of the
 helicity conserving amplitudes; while the fermion and Higgs
contributions are very small there. For the real parts of these
amplitudes  though, as well for the other (small) helicity amplitudes,
the $W$ contributions are at the same level as the other ones
in SM; their sum being always very small.
Similar results have also been observed for the
$\gamma \gamma \to \gamma \gamma $ \cite{GPRgamma}
and $\gamma \gamma \to \gamma Z$ \cite{gggZ} cases;
but in these cases the asymptotic region starts already at
$\sim250~GeV$.\par

To assess the quality of the SM asymptotic expressions
given in Appendix B, we have
compared them to the exact 1-loop results for the ten
$\gamma \gamma \to ZZ$ amplitudes in (\ref{8basictext},
\ref{-beta++--text} \ref{-beta++-0text}).
We find that at $\sqrt{\s} \simeq 1~ TeV$,
the differences between the   imaginary parts of the
asymptotic and  exact 1-loop results,
are at the $10\%$ level or smaller. At higher energies the agreement
improves of course,  reaching the level of the fourth significant digit
at $\sim 10~TeV $. For the  other  amplitudes though,
almost complete cancellations among the
various  terms occur, particularly for $\s \gsim 1~ TeV^2$;
leading to the conclusion,  (common for both
the asymptotic and the exact 1-loop expressions),
that they  are indeed   negligible.
\par

\vspace{0.5cm}
We next turn to the possible SUSY contributions  to
the various amplitudes.
As such we study contributions from a chargino  or a sfermion loop,
either in  diagrams with four external legs, or in Higgs-pole
diagrams involving a  Higgs-$\gamma \gamma $ vertex.\par

\vspace{0.4cm}
{\bf The chargino contribution.}\\
The contribution from the lightest positively charged
 chargino  $\tilde \chi^+_1$
is obtained from the effective interaction (\ref{LZff}) by using
\cite{ SUSY-rev, abdel9806}
\bqa
g^Z_{v\tilde \chi_1}& = &{1\over 2c_Ws_W}
\left \{ {3\over2}-2s^2_W +{1\over4}
[\cos(2\phi_L)+\cos(2\phi_R)]\right \}  \ , \label{gvchi1Z} \\
g^Z_{a\tilde \chi_1} & =& \frac{1}{8\cw\sw} \left [
\cos(2\phi_R) -\cos(2\phi_L) \right ] ~ , \label{gachi1Z}
\eqa
with
\bqa
\cos(2\phi_L)&=&-{M^2_2-\mu^2-2\mwd \cos(2\beta)\over
\sqrt{(M^2_2+\mu^2+2\mwd)^2-4[M_2\mu-\mwd \sin(2\beta)]^2}}
\ ,\nonumber\\
\cos(2\phi_R)&=&-{M^2_2-\mu^2+2\mwd \cos(2\beta)\over
\sqrt{(M^2_2+\mu^2+2\mwd)^2-4[M_2\mu-\mwd \sin(2\beta)]^2}}
\ ,
\eqa
and
\bq
M^2_{\tilde \chi_1^+}={1\over2} \left \{M^2_2+\mu^2+2\mwd -
\sqrt{(M^2_2+\mu^2+2\mwd )^2-4[M_2\mu-\mwd \sin(2\beta)]^2}~
\right \}, \label{chimass}
\eq
where $M_2$ and $\mu$ are taken real, and $\beta$ is the usual SUSY
parameter. These formulae should be combined with
(\ref{Ffamp}, \ref{deltafv} -\ref{fa+-+-})
in Appendix A, in order to calculate the chargino loop contribution
to the four-leg diagrams.\par

In  SUSY, the Higgs-pole contribution,
due to the lightest chargino
$\tilde \chi_1^+$ loop affecting the Higgs-$\gamma \gamma$ vertex,
may in general involve any of the two CP-even neutral Higgs states
$h^0$ or $H^0$.
Since we will be working below in the so called decoupling regime,
where $m_A \sim m_{H^0} \sim m_{H^\pm} \gg \mz$, we only need
the $h^0ZZ$ and $h^0 \chi_1^+\chi_1^-$ interaction lagrangian
\cite{SUSY-rev}
\bqa
 \L_{h^0ZZ,h^0 \chi_1^+\chi_1^-}& = &
\frac{g \mz}{2\cw}\,  \sin(\beta-\alpha) h^0 Z_\mu Z^\mu
\nonumber \\
& - & \frac{g}{\sqrt{2}}
[-\sin \alpha \cos \phi_R \sin\phi_L +
\cos \alpha \cos \phi_L \sin \phi_R ] h^0 \bar{\tilde \chi}_1^+
\tilde \chi_1^+  ~ . \label{h-chargino}
\eqa
Comparing this with (\ref{h-int})  and working in the
decoupling SUSY regime where $\alpha=\beta-\pi/2$, we
write the lightest chargino contribution to the curly brackets
in (\ref{Fh}) as
\bq
\H_{\tilde \chi_1^+}(\tau_{\tilde \chi_1^+})=
\frac{\sqrt{2}\mw}{m_{\tilde \chi_1}}
[ \cos \beta \cos \phi_R \sin\phi_L +
\sin \beta \cos \phi_L \sin \phi_R ]\, F_{1/2}(\tau_{\tilde
\chi_1^+}) ~ , \label{Hchi}
\eq
where
\bq
\tau_{\tilde \chi_1^+} \equiv \frac{4 m^2_{\tilde \chi_1^+}}{\s}
~ , \label{tauchi}
\eq
and  (\ref{F12}, \ref{ftau}) should be used.\par

\begin{figure}[p]
\vspace*{-4cm}
\[
\epsfig{file=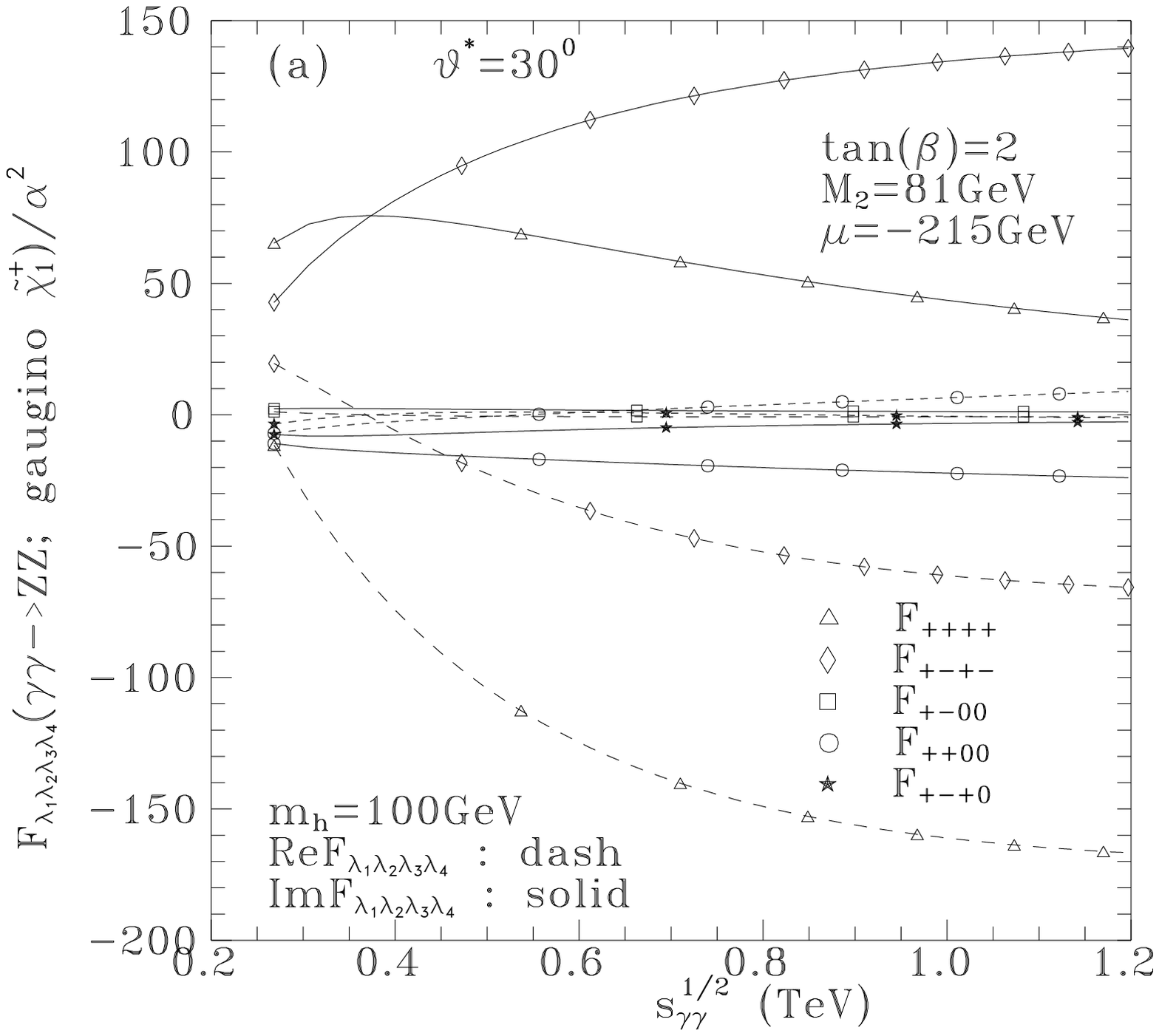,height=7.5cm}\hspace{0.5cm}
\epsfig{file=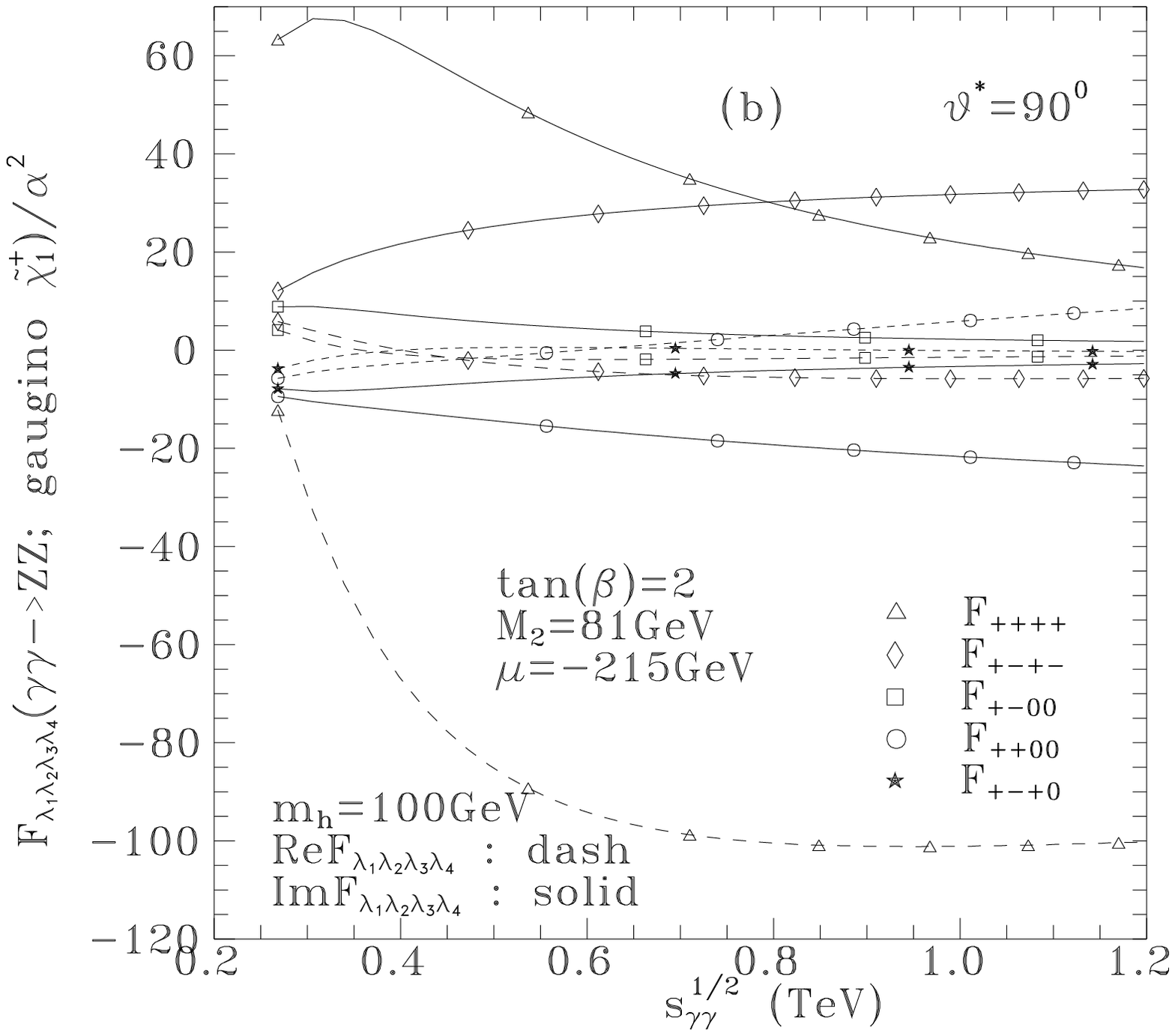,height=7.5cm}
\]
\vspace*{0.5cm}
\[
\epsfig{file=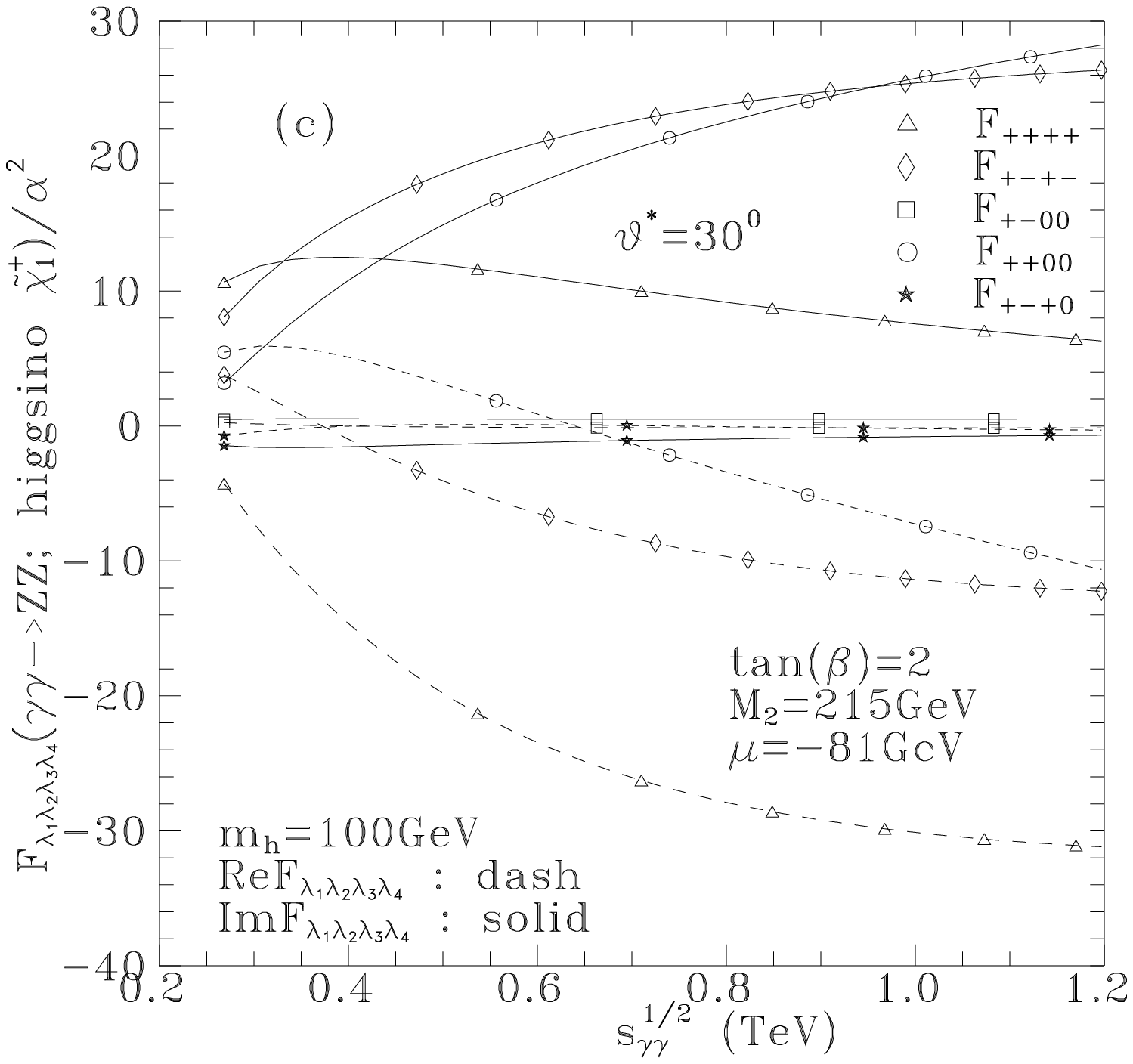,height=7.5cm}\hspace{0.5cm}
\epsfig{file=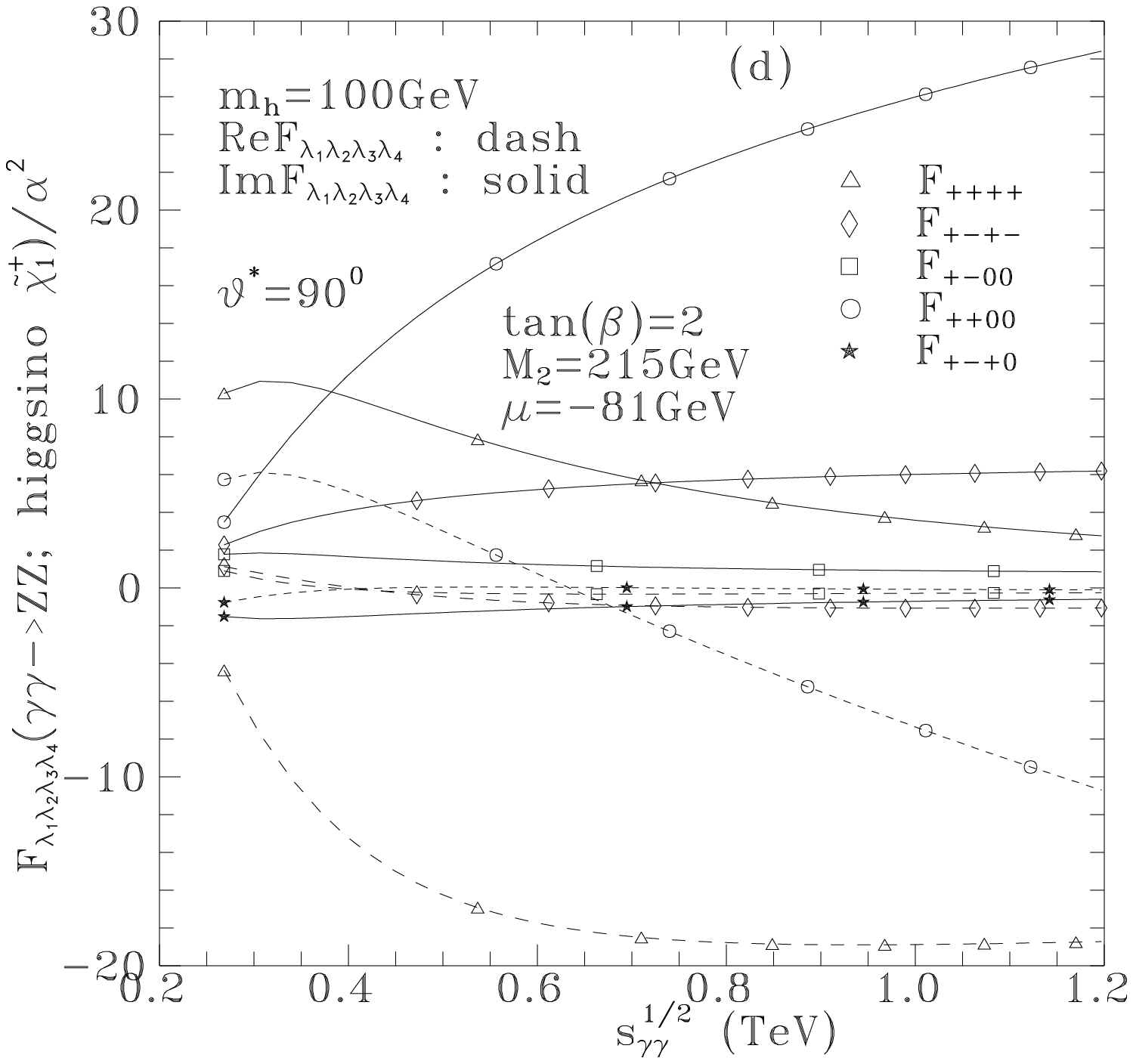,height=7.5cm}
\]
\vspace*{0.5cm}
\caption[1]{Chargino contribution to $\gamma \gamma \to Z Z$
helicity amplitudes for the gaugino  and higgsino
cases  at $\vartheta^*=30^0$ and $\vartheta^*=90^0$. The
parameters used are indicated in the figures
and $Q_{\chi^+_1}=1$.}
\label{chargino-amp}
\end{figure}

Using the relations (\ref{gvchi1Z}-\ref{tauchi}),
together  with the results (\ref{Fh}, \ref{Ffamp},
\ref{deltafv}-\ref{fa+-+-}) of Appendix A, and the exact
calculation  of the 1-loop functions provided by
 \cite{Oldenborgh}, we present in Fig.\ref{chargino-amp} the results for
two almost "extreme" situations corresponding
to a light chargino of mass $M_{\tilde \chi_1^+}\simeq
95~GeV$, with $\tan\beta=2$ and $\mu< 0$.
In the first case the chargino nature
is taken gaugino-like, by choosing (see Fig.\ref{chargino-amp}a,b)
\bqa
  M_2=81~ GeV  & , &   \mu=-215~ GeV ~~~~,
\nonumber  \\
 g^Z_{v\tilde \chi_1}= 1.72 & , &
g^Z_{a\tilde \chi_1}=0.102  ~;    \label{gaugino-par}
\eqa
while in the second case it is taken "higgsino-like" by choosing
(see Fig.\ref{chargino-amp}c,d)
\bqa
 M_2=215 ~GeV & , & \mu=-81~ GeV ~~~~~
\nonumber \\
g^Z_{v\tilde \chi_1}= 0.76  & , & ~~~
g^Z_{a\tilde \chi_1}= 0.113 \ . \label{higgsino-par}
\eqa
The conclusion from Fig.\ref{chargino-amp} is that
 $\gamma \gamma \to ZZ$  is
much more sensitive to a gaugino-like chargino, than to a
higgsino-like. This fact was also observed in the $\gamma \gamma
\to \gamma Z$ case; while $\gamma \gamma \to \gamma \gamma $
 is of course equally sensitive to both.
The Higgs-pole contribution turns out to be quite
small in  the chargino case; so  that the main effect arises from the
chargino loop in the four-external-leg diagrams.
Similar results, would also be obtained if the gaugino-like
state would correspond to a $\mu>0 $  solution,
like \eg\@   $M_{\tilde \chi_1^+}\simeq 96~ GeV$,  $\tan\beta=2.5$,
$M_2=120~ GeV$ and $\mu=300~ GeV$  \cite{abdel9907}. \par

\vspace{0.4cm}
{\bf The contributions from  a slepton or the
lightest stop $\tilde t_1$}\\
As in the chargino case, we consider the decoupling limit
 $\alpha=\beta-\pi/2$ for the   charged slepton  and the
 lightest stop contributions. Then, the mass-terms and the
$h^0\tilde e^*_{L(R)}\tilde e_{L(R)}$ and $h^0 \tilde t_1^* \tilde t_1$
interaction  Lagrangian are given by  \cite{SUSY-rev, Gunion-book}
\bqa
\L_{h^0\tilde f \tilde f} & = & -
\left ( \matrix{\tilde t_L^* & \tilde t_R^* } \right )
\left (
\matrix {
   M_{\tilde tL}^2 +m_t^2 & m_t \tilde A_t \cr
  m_t \tilde A_t &  M_{\tilde tR}^2 +m_t^2 } \right )
 \left ( \matrix{\tilde t_L \cr \tilde t_R } \right )
 - M_{\tilde eL}^2 \tilde e_L^*\tilde e_L
- M_{\tilde eR}^2 \tilde e_R^*\tilde e_R
\nonumber \\
&& -\, \frac{g \mzd }{\mw} \cos(2\beta)h^0 \Big [
\Big (\frac{1}{2}- \frac{2}{3} \swd \Big )\tilde t_L^*\tilde t_L
+\frac{2\swd}{3}\tilde t_R^*\tilde t_R
+ \Big (-\frac{1}{2} + \swd \Big )\tilde e_L^*\tilde e_L
-\swd \tilde e_R^*\tilde e_R \Big ]
\nonumber \\
&& -\, \frac{g m_t\tilde A_t}{2 \mw}\, h^0 (\tilde t_L^*\tilde t_R +
\tilde t_R^*\tilde t_L)
-\, \frac{g m_t^2}{\mw}\, h^0 (\tilde t_L^*\tilde t_L +
\tilde t_R^*\tilde t_R) ~ , \label{h-sfermion}
\eqa
where
\bq
\tilde A_t = A_t -\mu \cot(\beta) ~ ~ , \label{At-tilde}
\eq
and
$ M_{\tilde tL}~,~  M_{\tilde tR}$,
$ M_{\tilde eL}~,~  M_{\tilde eR}$, $A_t$ are the
usual  soft breaking parameters in the stop and slepton sector
\cite{SUSY-rev, Gunion-book}. Eqs.  (\ref{h-sfermion},
\ref{At-tilde}) determine the sfermion Higgs-pole contributions
and possible mixing; while  the loop contributions
due to a scalar particle  with definite weak isospin and charge,
  are given by (\ref{FSamp}-\ref{S+-+-}).\par

We first discuss the charged  slepton case for which
there is no appreciable mixing, so that we are lead
to   a pure \eg\@ L- or  R-selectron
circulating along the loop; compare (\ref{h-sfermion}).
Taking  a common mass
$M_{\tilde e}= M_{\tilde eL} =M_{\tilde e R}= 0.1~TeV$, for both
($\tilde e_L$ ,   $\tilde e_R $) in (\ref{h-sfermion}); we get
for  a selectron loop  with  definite isospin
and charge
\bq
g^Z_{\tilde e}={1\over c_Ws_W}[t^{\tilde e}_3 - Q_{\tilde e} s^2_W]
\ \ , \label{gsleptonZ}
\eq
to be used in (\ref{FSamp} - \ref{S+-+-}) in Appendix A,
with   $Q_{\tilde e_L}=Q_{\tilde e_R}=-1$,
$t^{\tilde e_L}_3=-{1\over2}$ and $t^{\tilde e_R}_3=0$;
compare (\ref{gSZ}).\par

 For an L-selectron this leads  to
$g^Z_{\tilde e_L}=-0.65$, while the Higgs-pole contribution
is obtained by comparing  (\ref{h-sfermion},
\ref{h-int}) to be
\bq
\H_{\tilde e_L}(\tau_{\tilde e})=
\frac{\mzd}{M_{\tilde e}^2} \cos(2\beta) (-\, \frac{1}{2} +\swd)
F_0(\tau_{\tilde e}) ~ , \label{HLselectron}
\eq
where
\bq
\tau_{\tilde e} = \frac{4 M_{\tilde e}^2}{\s} ~ .
\label{tauselectron}
\eq \par

 Correspondingly, for an R-selectron, we have
 $g^Z_{\tilde e_R}=+ 0.54$, while
the Higgs-pole contribution is determined by
\bq
\H_{\tilde e_R}(\tau_{\tilde e})=
\frac{\mzd}{M_{\tilde e}^2} \cos(2\beta) (-\swd)
F_0(\tau_{\tilde e}) ~ . \label{HRselectron}
\eq

\begin{figure}[p]
\vspace*{-4cm}
\[
\epsfig{file=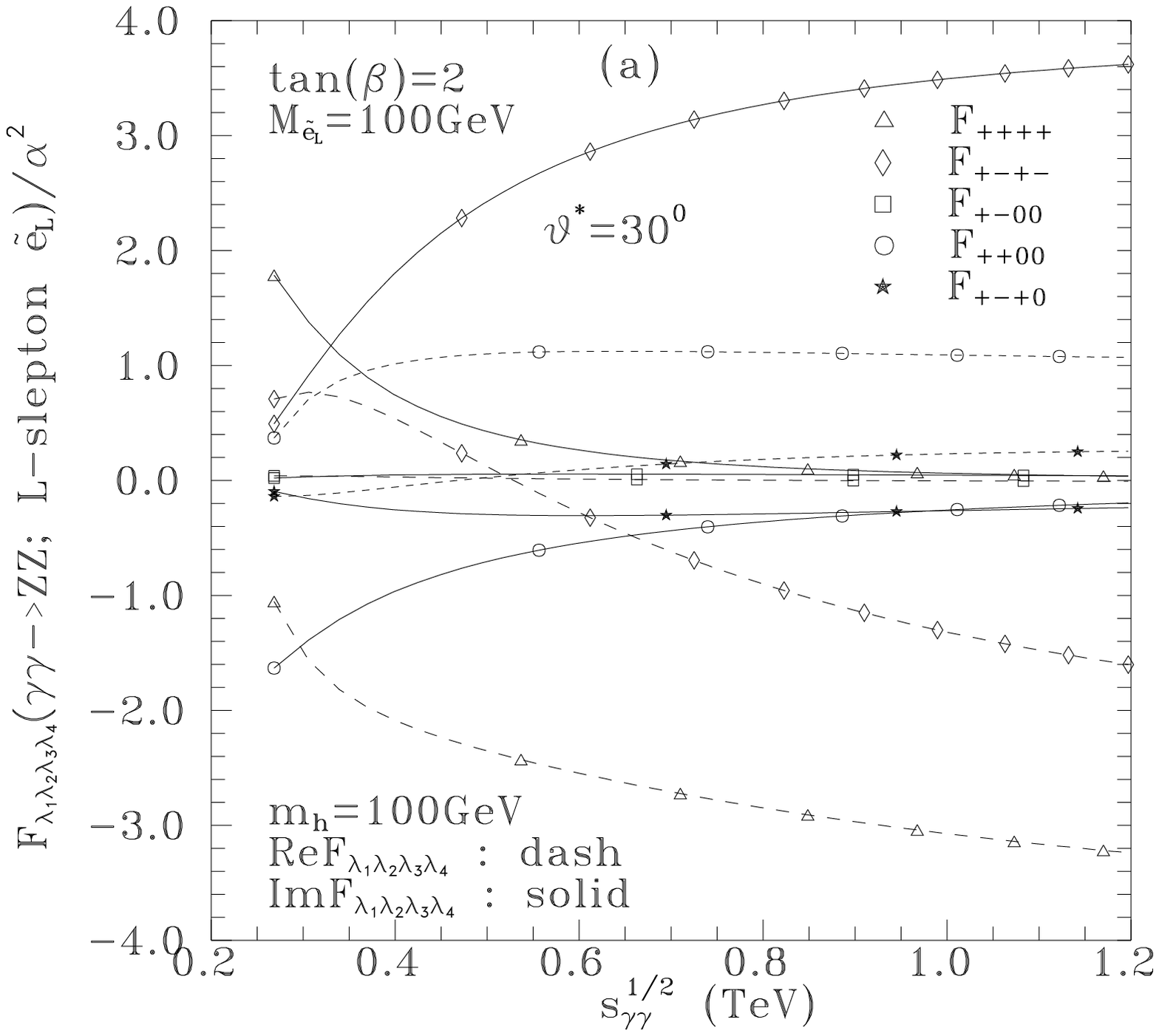,height=7.5cm}\hspace{0.5cm}
\epsfig{file=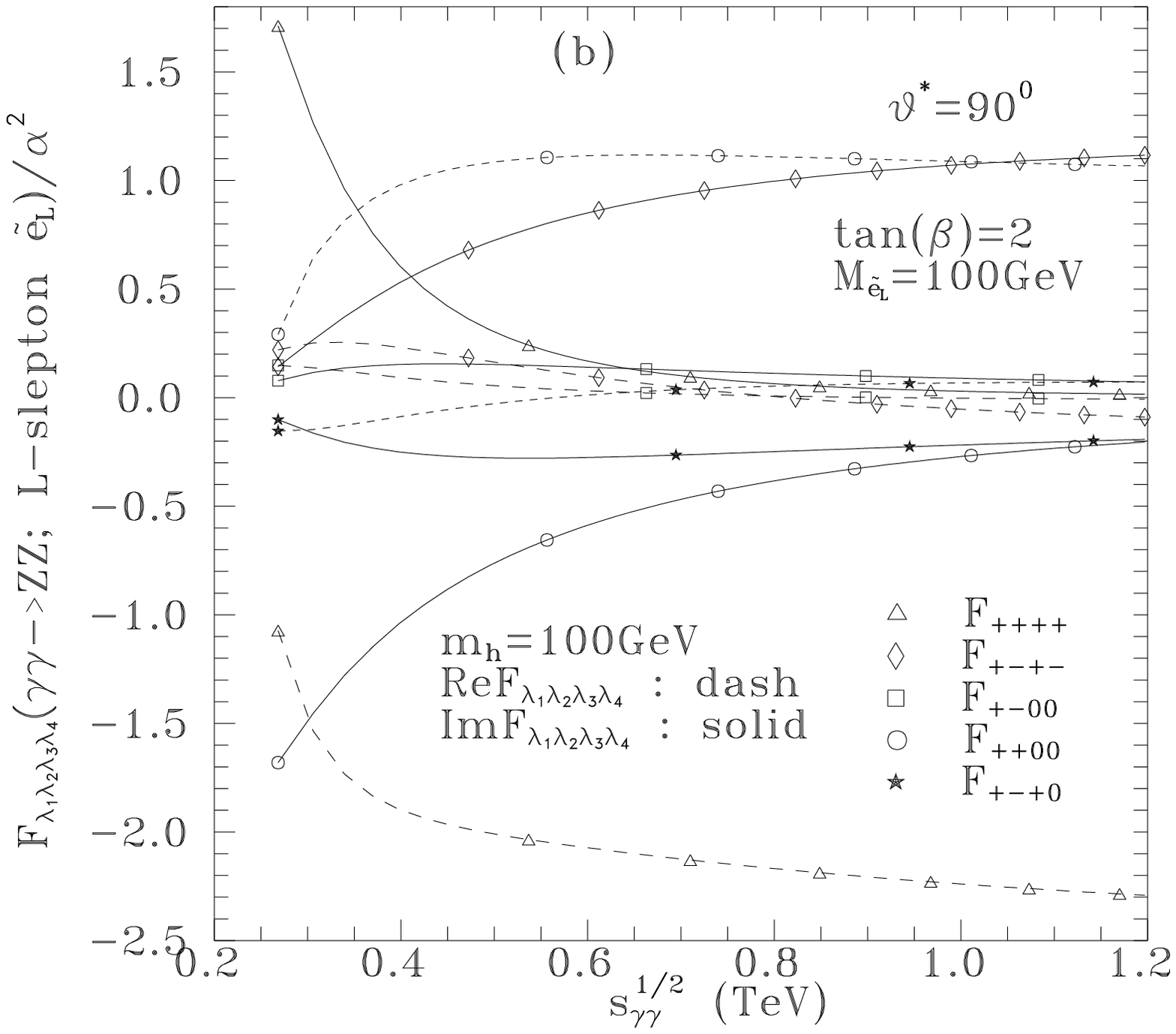,height=7.5cm}
\]
\vspace*{0.5cm}
\[
\epsfig{file=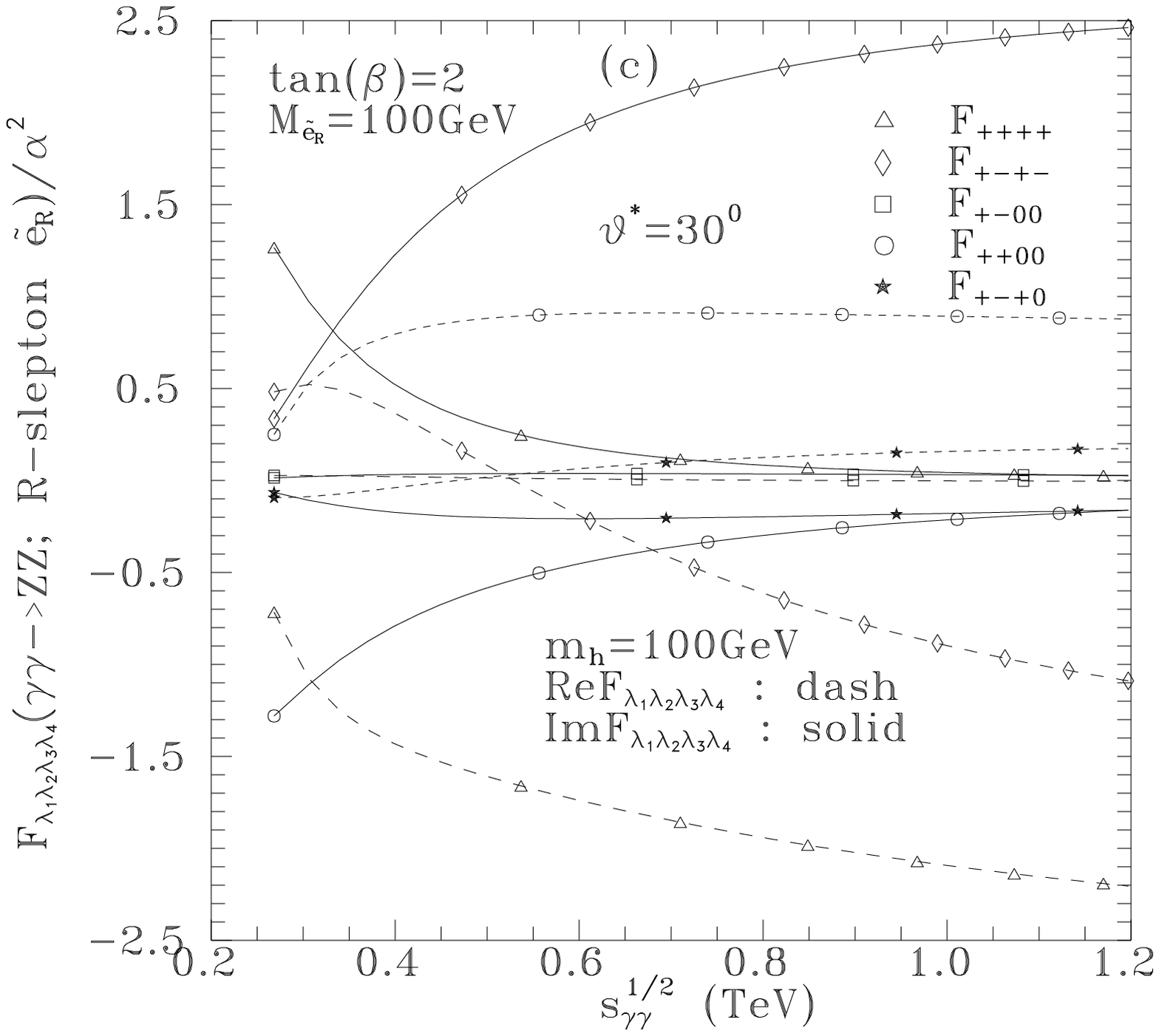,height=7.5cm}\hspace{0.5cm}
\epsfig{file=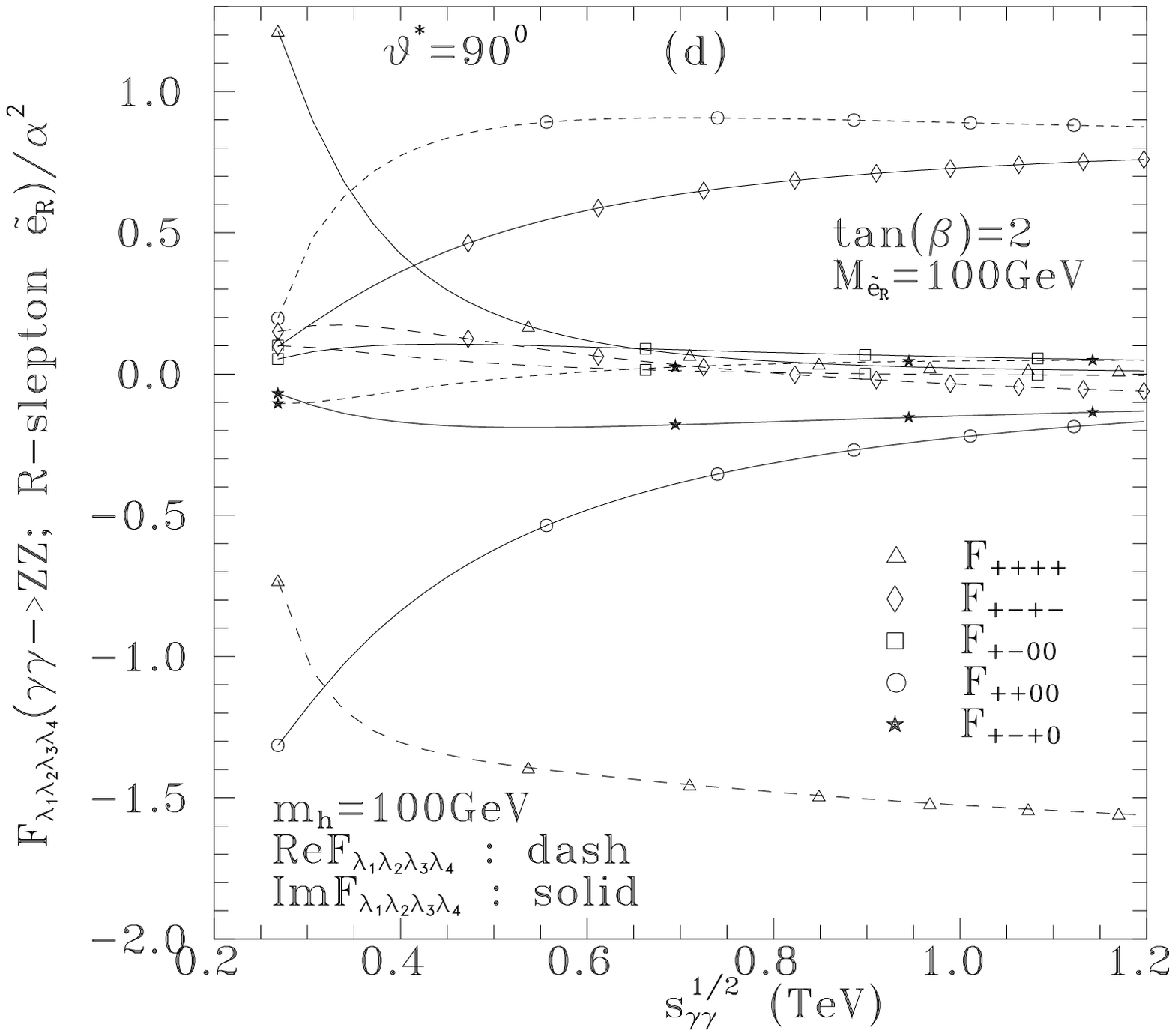,height=7.5cm}
\]
\vspace*{0.5cm}
\caption[1]{Contribution to $\gamma \gamma \to Z Z$
helicity amplitudes from a $\tilde e^-_L$ or $\tilde e^-_R$
loop, at $\vartheta^*=30^0$ and $\vartheta^*=90^0$.
The parameters used are indicated in the figures and the slepton mass
is taken $M_{\tilde e}=100GeV$.}
\label{slepton-amp}
\end{figure}

Substituting in   (\ref{Fh}, \ref{FSamp}),
 we find  that the   R- and L-selectrons give
very similar contributions to the $\gamma \gamma \to ZZ$
amplitudes; which is  confirmed by the results in
Fig.\ref{slepton-amp}a-d, derived using  the exact 1-loop
functions  in ( \ref{S+++-} - \ref{S+-+-}).
We recall that the R- and L-selectrons  contribute in the same way
also in the $\gamma \gamma \to \gamma \gamma $ amplitudes,
while their contributions to $\gamma \gamma \to \gamma Z$
have opposite signs \cite{gggg, gggZ}. It  seems
that $\gamma \gamma \to \gamma \gamma $ is somewhat more
sensitive to slepton contributions, than the other two processes
$\gamma \gamma \to \gamma Z ~, ~ZZ$.
It is also found that the slepton contributions  to $F_{++++}$ and
 $F_{++00}$,  due to the Higgs-pole or the  four-leg loop
diagrams, are comparable. \par

We next turn to the contribution from the lightest stop,
denoted as $\tilde t_1$, which is  obtained by taking into
account the mixing
implied by (\ref{h-sfermion}). For real $M_{\tilde tL}$,
$M_{\tilde tR}$ and $\tilde A_t$, this leads to
\bq
\left (\matrix{\tilde t_L \cr \tilde t_R} \right )=
\left ( \matrix {
  \cos \theta_t &  -\sin \theta_t \cr
 \sin \theta_t &  \cos \theta_t } \right )
\left (\matrix{\tilde t_1 \cr \tilde t_2} \right )
~ \label{stop-mixing}
\eq
\bq
m^2_{\tilde t_1, \tilde t_2}  =
\frac{1}{2}  \Big \{ M^2_{\tilde tL} + M^2_{\tilde tR} +2 m_t^2
\mp \sqrt{(M^2_{\tilde tL} - M^2_{\tilde tR})^2 +4 m_t^2 \tilde A_t^2 }
~ \Big \} ~ , \label{stop1mass}
\eq
\bqa
\sin (2\theta_t)=\frac{2m_t \tilde A_t}{m^2_{\tilde t_1}
-m^2_{\tilde t_2}} & , &
\cos({2 \theta_t}) = \frac {M^2_{\tilde tL} - M^2_{\tilde tR}}
{m^2_{\tilde t_1} -m^2_{\tilde t_2}} ~.
\label{thetat}
\eqa
Then, the $Z$-stop coupling to be
used in  conjunction with (\ref{FSamp}) is
\bq
g^Z_{\tilde t_1} = \frac{1}{2\cw \sw} \Big
(  \cos^2 \theta_t -\frac{4}{3} \swd \Big )
~, \label{gZstop1}
\eq
while (\ref{h-sfermion}, \ref{h-int}, \ref{Fh} ) determine the
$\tilde t_1$ Higgs-pole contribution by
\bq
\H_{\tilde t_1}(\tau_{\tilde t_1}) =\frac{3}{m_{\tilde t_1}^2}
\Bigg \{ \mzd \cos(2\beta) \Big [ \frac{\cos^2\theta_t}{2}
-\, \frac{2\swd}{3} \cos(2\theta_t)  \Big ]
+\frac{m_t \tilde A_t}{2} \sin(2\theta_t) +m_t^2 \Bigg \}
F_0(\tau_{\tilde t_1})     ~ , \label{Hstop1}
\eq
where
\bq
\tau_{\tilde t_1}=\frac{4 m^2_{\tilde t_1}}{\s} ~ ,
\label{taustop1}
\eq
and the factor three for colour multiplicity has been
included.\par

\begin{figure}[p]
\vspace*{-4cm}
\[
\epsfig{file=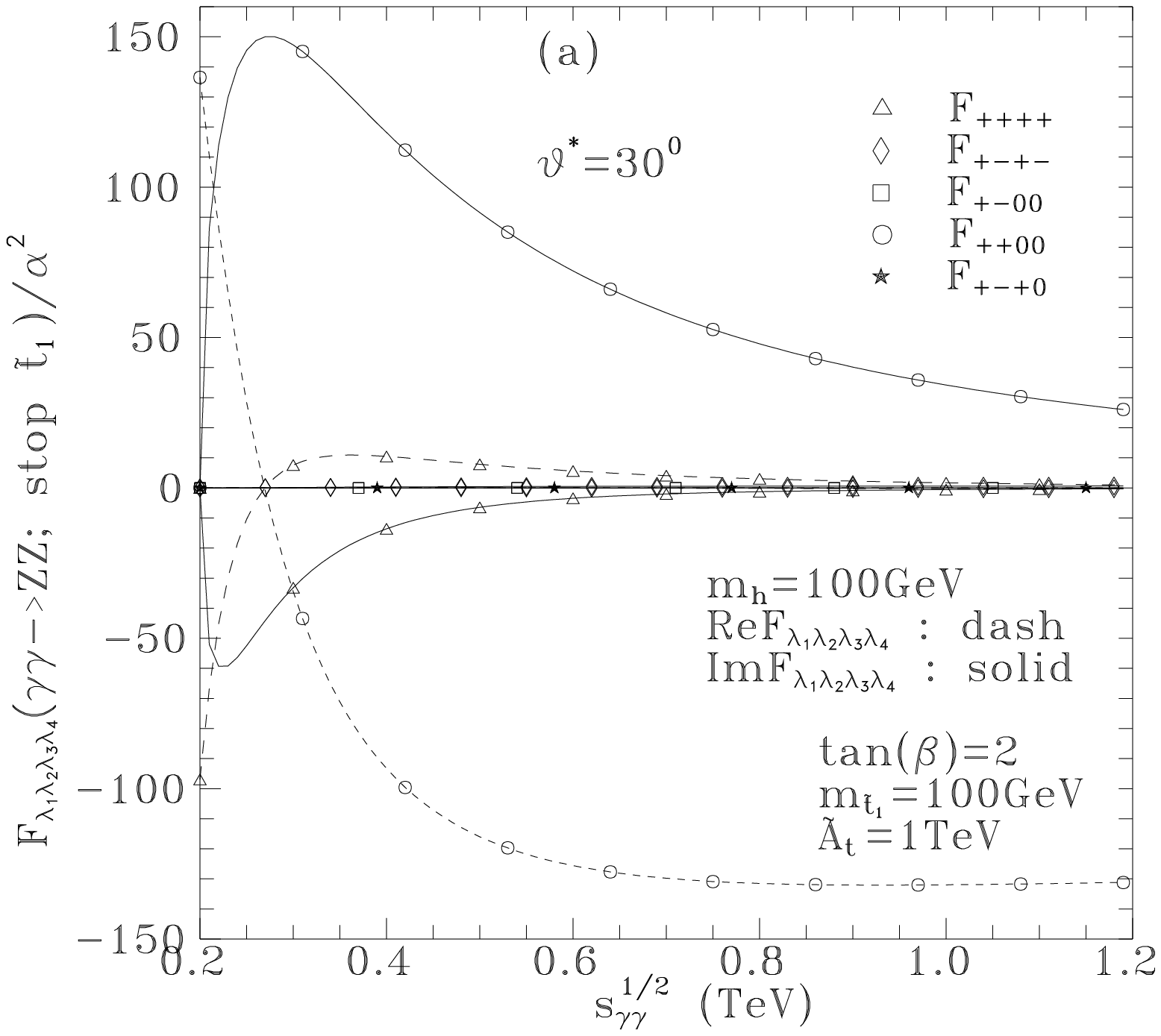,height=7.5cm}\hspace{0.5cm}
\epsfig{file=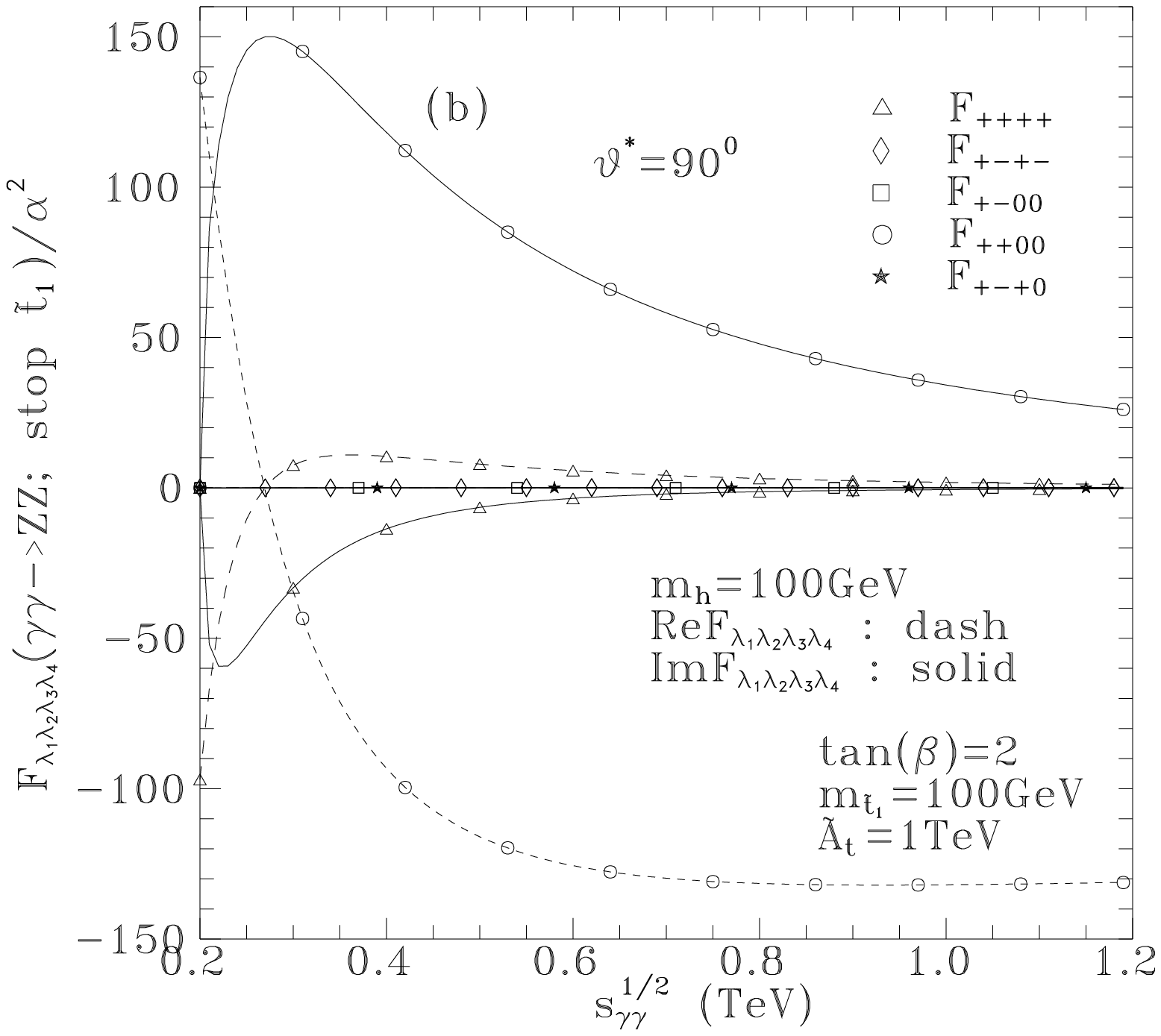,height=7.5cm}
\]
\caption[1]{Contribution to $\gamma \gamma \to Z Z$
helicity amplitudes from the lightest stop  $\tilde t_1$
at $\vartheta^*=30^0$ and $\vartheta^*=90^0$.
The parameters used are indicated in the figures.}
\label{stop1-amp}
\end{figure}

An example of a lightest stop contribution to the
$\gamma \gamma \to ZZ $ amplitudes is given in
Fig.\ref{stop1-amp},  corresponding to the assumption that
$M_{\tilde tL}= M_{\tilde tR}$ are chosen so  that
$m_{\tilde t_1}=100~ GeV$, and $\tilde A_t=1~ TeV$.
In such a case we get  $\theta_t=3 \pi/4$.
As shown in Fig.\ref{stop1-amp}, the $\tilde t_1$
contributions to the
amplitudes, are almost independent of $\vartheta^*$;
which simply indicates the dominance
of the Higgs-pole contribution.\par

\vspace{0.5cm}
A comparison of Fig.\ref{chargino-amp}a-d,
 Fig.\ref{slepton-amp}a-d and Fig.\ref{stop1-amp}a,b
indicates that the most promising effects are  generated
either by a gaugino-like chargino, or from the lightest stop  $\tilde t_1$.
Most of the $\tilde t_1$ effect arises from the Higgs-pole
contribution to the  amplitudes. This explains why the stop effect
is much smaller in the $\gamma \gamma \to \gamma \gamma $, $\gamma Z$
cases \cite{gggg, gggZ}, where this last contribution is absent. \par

\section{ The $\gamma \gamma \to Z Z$ Cross sections}

\hspace{0.7cm}We next explore
the possibility to use polarized or unpolarized
$\gamma\gamma$ collisions in an $LC_{\gamma \gamma}$ Collider
\cite{ GPRgamma}.  As in the $\gamma \gamma \to
\gamma \gamma $ case \cite{gggg}, Bose statistics and Parity
invariance leads to
\bqa
{d\sigma\over d\tau d\cos\vartheta^*}&=&{d \bar L_{\gamma\gamma}\over
d\tau} \Bigg \{
{d\bar{\sigma}_0\over d\cos\vartheta^*}
+\langle \xi_2 \xi_2^\prime
\rangle{d\bar{\sigma}_{22}\over d\cos\vartheta^*}
+\left [ \langle\xi_3\rangle \cos2\phi
+\langle\xi_3^ \prime\rangle\cos2\phi^\prime \right ]
{d\bar\sigma_3 \over d\cos\vartheta^*}
\nonumber\\
&&+\langle\xi_3 \xi_3^\prime\rangle
\left[{d\bar{\sigma}_{33}\over d\cos\vartheta^*}
\cos2(\phi+\phi^\prime)
+{d\bar{\sigma}^\prime_{33}\over
d\cos\vartheta^*}\cos2(\phi- \phi^\prime)\right ]\nonumber\\
&&+ \left [ \langle\xi_2 \xi_3^\prime\rangle\sin2 \phi^\prime
- \langle\xi_3 \xi_2^\prime\rangle\sin2\phi \right ]
{d\bar{\sigma}_{23}\over d\cos\vartheta^*} \Bigg \} \ \ ,
\label{sigpol}
\eqa
where
\bqa
{d\bar \sigma_0(\gamma \gamma \to ZZ) \over d\cos\vartheta^*}&=&
\left ({\beta_Z\over 128 \pi\hat{s}}\right )
\sum_{\lambda_3\lambda_4} [|F_{++\lambda_3\lambda_4}|^2
+|F_{+-\lambda_3\lambda_4}|^2] ~ ,  \label{sig0} \\
{d\bar{\sigma}_{22}(\gamma \gamma \to ZZ)\over d\cos\vartheta^*} &=&
\left ({\beta_Z\over 128 \pi\hat{s}}\right )\sum_{\lambda_3\lambda_4}
[|F_{++\lambda_3\lambda_4}|^2
-|F_{+-\lambda_3\lambda_4}|^2]  \ , \label{sig22} \\
{d\bar{\sigma}_{3}(\gamma \gamma \to ZZ)\over d\cos\vartheta^*}&=&
\left ({-\beta_Z\over 64 \pi\hat{s}}\right ) \sum_{\lambda_3\lambda_4}
Re[F_{++\lambda_3\lambda_4}F^*_{-+\lambda_3\lambda_4}]  \ ,
\label{sig3} \\
{d\bar \sigma_{33}(\gamma \gamma \to ZZ) \over d\cos\vartheta^*}& = &
\left ({\beta_Z\over 128\pi\hat{s}}\right ) \sum_{\lambda_3\lambda_4}
Re[F_{+-\lambda_3\lambda_4}F^*_{-+\lambda_3\lambda_4}] \ ,
\label{sig33} \\
{d\bar{\sigma}^\prime_{33}(\gamma \gamma \to ZZ)\over d\cos\vartheta^*}&=&
\left ({\beta_Z\over 128 \pi\hat{s}}\right ) \sum_{\lambda_3\lambda_4}
Re[F_{++\lambda_3\lambda_4}F^*_{--\lambda_3\lambda_4}] \  ,
\label{sig33prime} \\
{d\bar{\sigma}_{23}(\gamma \gamma \to ZZ)\over d\cos\vartheta^*}& = &
\left ({\beta_Z\over 64 \pi\hat{s}}\right ) \sum_{\lambda_3\lambda_4}
Im[F_{++\lambda_3\lambda_4}F^*_{+-\lambda_3\lambda_4}] \ ,
\label{sig23}
\eqa
are expressed in terms of the amplitudes given in Appendix A.
The quantity $d\bar L_{\gamma\gamma}/d\tau$
in  (\ref{sigpol}), describes the
photon-photon luminosity
per unit $e^-e^+$ flux, in an LC operated in the $\gamma \gamma$ mode
\cite{LCgg}.
The Stokes parameters $\xi_2$, $\xi_3$ and the azimuthal angle
$\phi$  in (\ref{sigpol}), determine the normalized
most general helicity density matrix
of one of the backscattered photons
$\rho^{BN}_{\lambda \tilde \lambda}$,
through the formalism described in Appendix B of \cite{gggg};
compare Eq.(B4) of \cite{gggg}.
The corresponding parameters for the other backscattered photon are
denoted by a prime.  The numerical expectations for
 $d\bar L_{\gamma\gamma}/d\tau$, $\langle \xi _j \rangle $,
$\langle \xi _j^\prime \rangle $ and
$\langle \xi_i \xi _j^\prime \rangle $ are given
in Appendix B and Fig.4 of \cite{gggg}. \par

In (\ref{sig0} - \ref{sig23}),
 $\beta_Z$ is the $Z$ velocity in the $ZZ $ frame,
while $\vartheta^*$ is the scattering angle, and
$\tau \equiv s_{\gamma \gamma}/s_{ee}$. Because of Bose
statistics, all $d\bar \sigma_j/ d\cos\vartheta^*$
are forward-backward symmetric.
Note that $d\bar \sigma_0/ d\cos\vartheta^*$ is the unpolarized
cross section. This is the only $\bar \sigma_j$ quantity which is
positive definite.\par

\begin{figure}[p]
\vspace*{-4cm}
\[
\epsfig{file=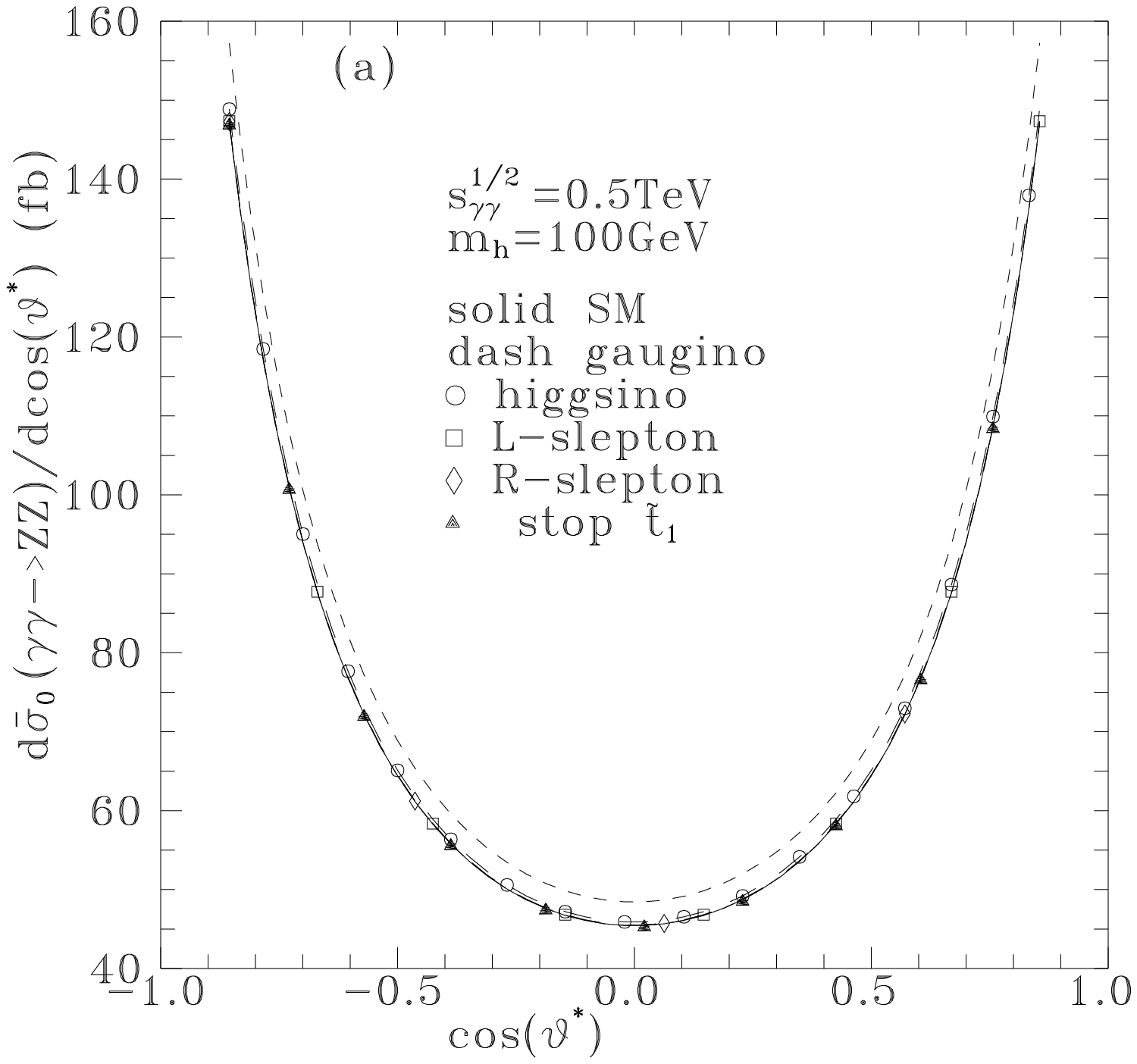,height=7.5cm}\hspace{0.5cm}
\epsfig{file=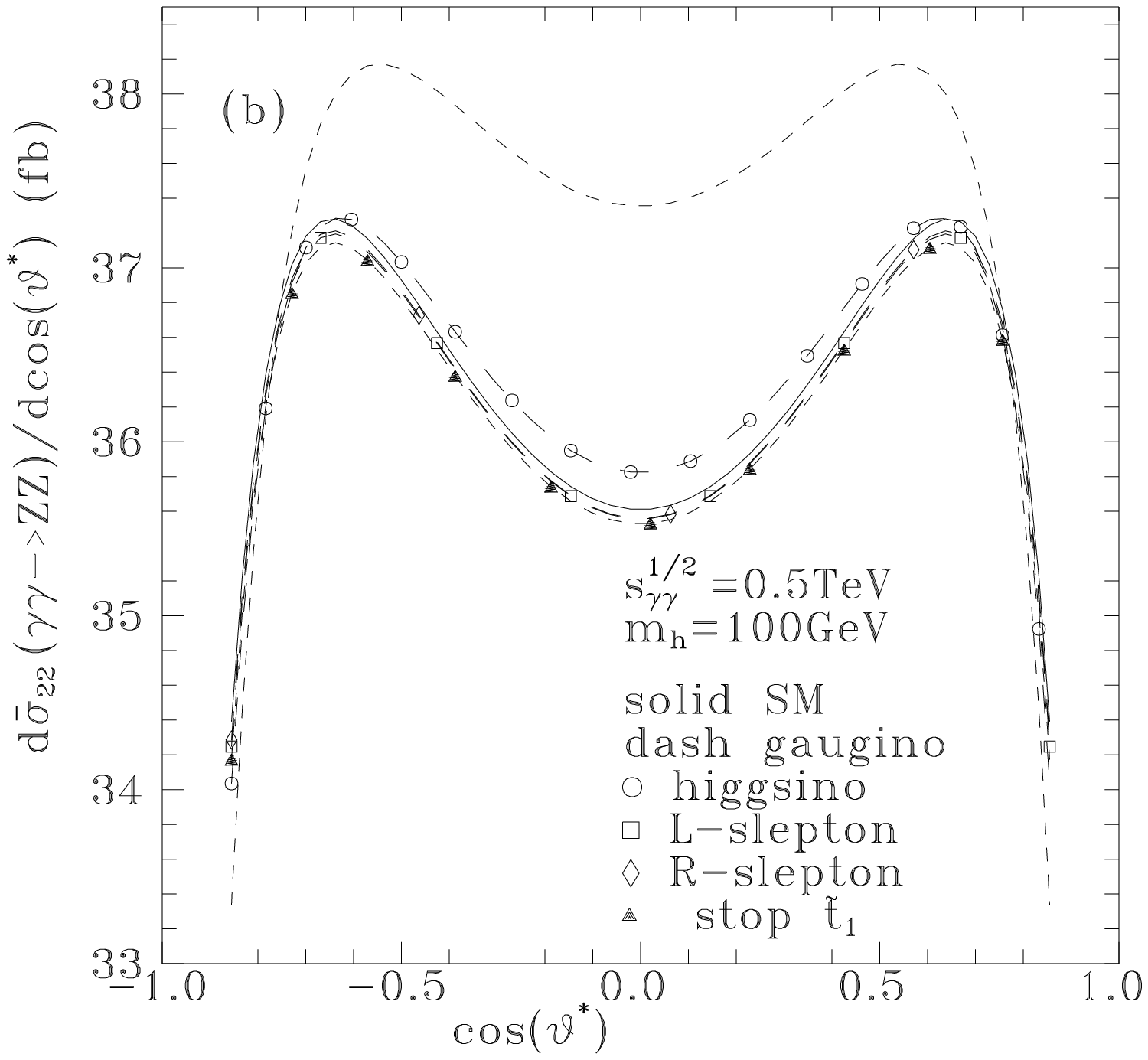,height=7.5cm}
\]
\vspace*{0.5cm}
\[
\epsfig{file=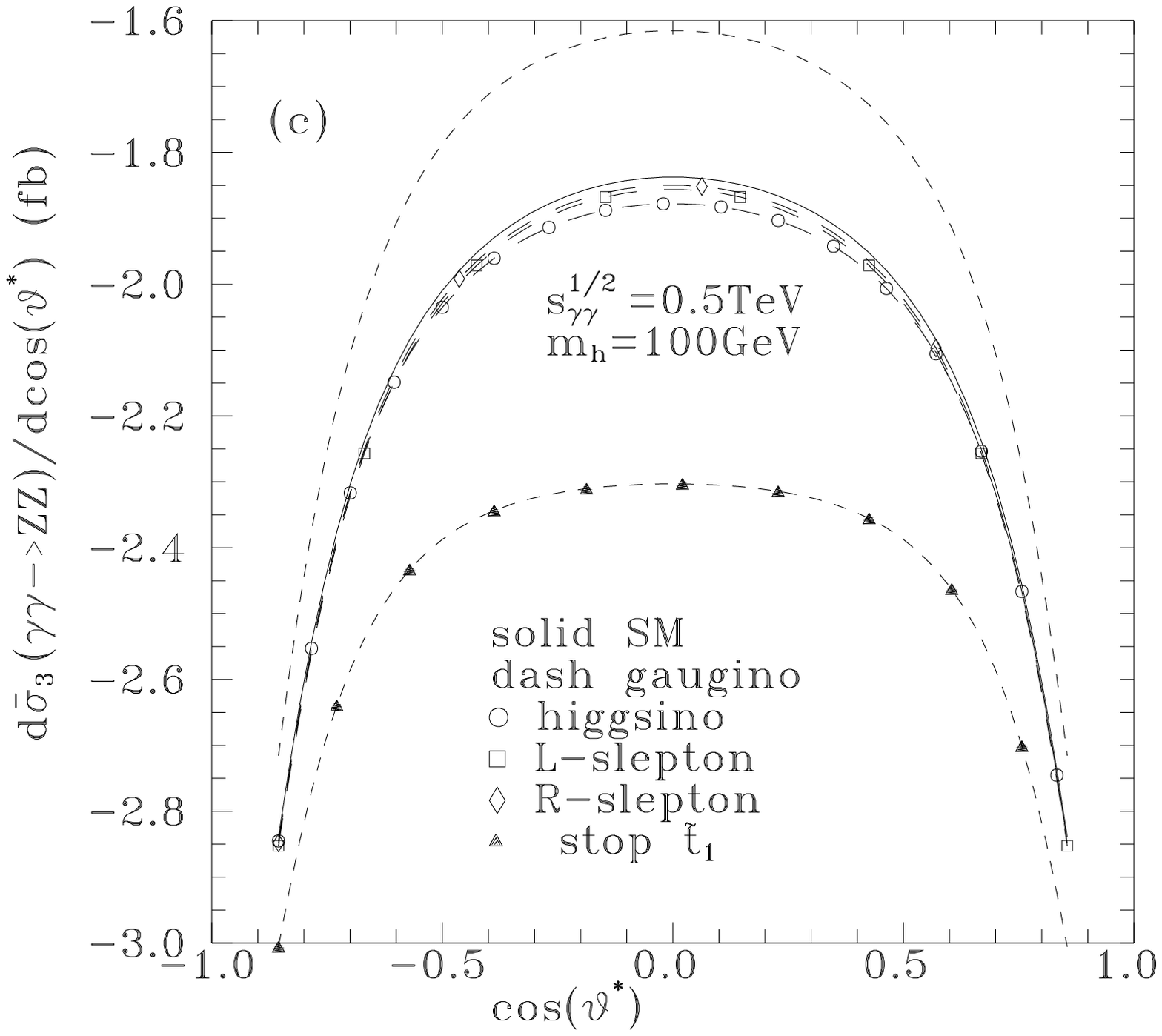,height=7.5cm}\hspace{0.5cm}
\epsfig{file=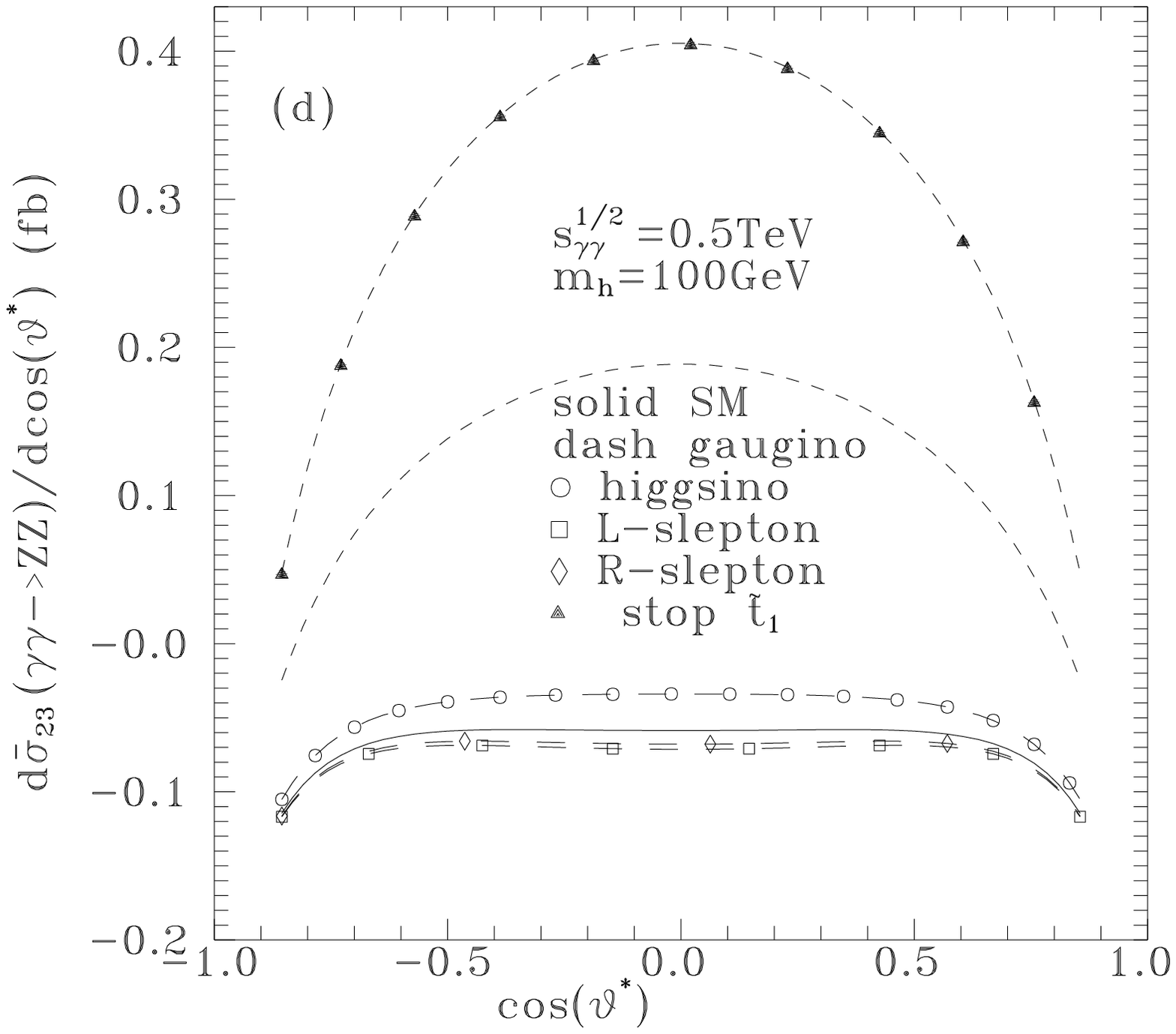,height=7.5cm}
\]
\vspace*{0.5cm}
\caption[1]{Angular distributions for
$d\bar{\sigma}_{0}/d\cos\vartheta^*$,
$d\bar{\sigma}_{22}/ d\cos\vartheta^*$,
$d\bar{\sigma}_{3}/d\cos\vartheta^*$,
$d\bar{\sigma}_{23}/d\cos\vartheta^*$.}
\label{angular}
\end{figure}

The results for the differential cross sections
$d\bar \sigma_j/d\cos\vartheta^* $, are given in
Fig.\ref{angular}a-f at $\sqrt {\s}=0.5~TeV$;
while the corresponding integrated cross sections in the
range $30^0 \leq \vartheta^* \leq 150^0$, appear
as functions of $\sqrt {\s}$, in Fig.\ref{sig}a-f.
In each case we give the standard model (SM)
predictions; as well as the results expected  for the   cases
of including the contributions from
a single chargino or a single charged slepton or the $\tilde t_1$.
For each of these SUSY contributions, we use the same parameters
as those appearing in the amplitudes presented in
Fig.\ref{chargino-amp}-\ref{stop1-amp}. \par

When comparing the general structure of the  differential
cross sections in Fig.\ref{angular}a-f, with the corresponding
results for $\gamma \gamma \to \gamma \gamma $ and $\gamma Z$
\cite{gggg, gggZ}, we  remark the following.
The general shape of $d\bar \sigma_0/d\cos\vartheta^* $ is roughly
the same in all three cases. Exactly the opposite shape, with central a
peak (at $\vartheta^*\simeq \pi/2$) and a dip in the forward and backward
regions, is found for  $d\bar \sigma_{22}/d\cos\vartheta^* $
in the $\gamma \gamma \to \gamma \gamma $ case; while for
$\gamma \gamma \to \gamma Z$ we find something  like a plateau
in the central region;
which develops to a central dip and two peaks at
$\vartheta^*\simeq \pi/4~,~3\pi/4$
for $\gamma \gamma \to ZZ$; compare Fig.6b of \cite{gggZ} and
Fig.\ref{angular}b of this paper.\par

The other cross sections are much smaller. Paying attention
only to the largest ones, we remark that
$d\bar \sigma_{33}/d\cos\vartheta^* $ has a central-peak
and a forward-backward dip structure for all processes;
compare Fig.6e in \cite{gggg} with Fig.\ref{angular}e here.
On the other hand, $d\bar \sigma_3/d\cos\vartheta^* $ has a
central plateau and  forward and  backward dips in
$\gamma \gamma \to ZZ$; which become a central plateau
accompanied with   forward and backward peaks in
$\gamma \gamma \to \gamma \gamma $; while in
$\gamma \gamma \to \gamma Z$ it is not forward-backward symmetric;
compare Fig.6d of \cite{gggZ} and Fig.\ref{angular}c.
\par

Concerning the \underline{relative} (NP versus SM) effects, the
main difference between  $\gamma \gamma \to ZZ$, and
$(\gamma \gamma \to \gamma \gamma ~,~ \gamma Z)$, is that the former
displays considerable sensitivity to the lightest stop
$\tilde t_1$, which is not shared by the other two.
This is because the lightest stop contribution is mainly
generated by the Higgs-pole diagrams; which of course
do not contribute to
$\gamma \gamma \to \gamma \gamma ~,~ \gamma Z$.
Such $\tilde t_1$ effects are mostly  visible
in $d\bar \sigma_{22}/d\cos\vartheta^* $ and
$d\bar \sigma_3/d\cos\vartheta^* $ shown in Fig.\ref{angular}b,c,
and in $\bar \sigma_{22}$ in Fig\ref{sig}c.
\par

With respect to the chargino signatures, the fact is   that
$\gamma \gamma \to ZZ$ and
$\gamma \gamma \to \gamma Z$ are mainly sensitive to a
gaugino-type chargino; while $\gamma \gamma \to \gamma \gamma$
is  equally sensitive to both, the gaugino- as well as the
higgsino-type charginos. Finally, very little sensitivity to
charged sleptons is displayed by all three processes
$\gamma \gamma \to \gamma \gamma ~,~ \gamma Z ~,~ ZZ$. \par

To make the discussion of  the observability  of the various
NP effects in the differential cross sections in
(\ref{sigpol}) more quantitative, we should  take into account
the experimental aspects of the $\gamma \gamma$ collision
realized through laser backscattering \cite{LCgg, gamma97}.
We proceed along the same lines as for the analysis of
the observable quantities  for $\gamma \gamma \to \gamma Z$
in Section 3 of \cite{gggZ}.
The differential cross sections for
$\gamma \gamma \to ZZ$ in Fig.\ref{angular}a-f, are
in almost all cases\footnote{The exception applies only to the
case of $d \bar \sigma_{23}/d\cos \vartheta^*$, which is
very small in SM, anyway.}  about a factor of 2 larger
than the corresponding
cross sections for $\gamma \gamma \to \gamma Z$  shown in
Fig.6a-f of \cite{gggZ}. Of course, for estimating the  number
of the measurable $ZZ$-production  events, some $ZZ$
identification factor should be taken into account.
A corresponding factor is
apparently not needed in the $\gamma Z$ production case, since
the photon provides a very good signature. Assuming that the
useful modes for the $ZZ$ identification are those where one $Z$
decays leptonically (including the invisible neutrino mode),
and the other hadronically, we get an identification factor of about
$1/2$; if only charged leptons are used for the leptonic modes this
factor decreases to 20 percent. So finally, the useful
$ZZ$ rate is comparable to the
$\gamma Z$ one. Thus, the statistical uncertainties in measuring the
various $ZZ$ cross sections are similar to those of the corresponding
$\gamma Z$ ones appearing in \cite{gggZ}.
Therefore, we expect that it
should be possible to attain an absolute accuracy of about $ 0.3fb$
for $ d \bar \sigma_0 (\gamma \gamma \to ZZ)/d \cos \vartheta^* $ at large
angles. Correspondingly, an absolute accuracy of about
$(0.3-3)fb $, (depending on the flux optimization),
should  be realistic   for the smaller quantities
$ d \bar \sigma_{22} /d \cos \vartheta^* $,
$ d \bar \sigma_3 /d \cos \vartheta^* $ and
$ d \bar \sigma_{33} /d \cos \vartheta^* $  at large angles.\par

Therefore, the $\gamma \gamma \to ZZ$  sensitivity to
a  gaugino-type chargino of $\sim100~ GeV$,
 is similar and even more pronounced then the sensitivity
of the $\gamma \gamma \to \gamma Z$ process;
while the higgsino or slepton effects are more depressed
in $\gamma \gamma \to ZZ$ \cite{gggZ}.
The important feature of the $ZZ$ production is its sensitivity to a
$\tilde t_1$ contribution, which may be comparable to the gaugino
or higgsino sensitivity, provided that sufficient
transverse and longitudinal
polarizations for the photon beams are available.
We also note that in the
present case we have explored this sensitivity only
in the decoupling limit.\par

The illustrations  given in the present paper are for a chargino,
slepton, or a lightest stop $\tilde t_1$ in 100~GeV mass range.
For higher masses, the relative merits of the
$\gamma\gamma\to ZZ$, $\gamma\gamma\to\gamma Z$  and
$\gamma\gamma\to\gamma\gamma$ processes\footnote{ In \cite{gggg}
we gave some illustration for sparticles at $250~GeV$ in the
$\gamma\gamma\to\gamma\gamma$ case.} remain about the same.
These processes should be very helpful in identifying the nature
of the various sparticles, up to masses of about 300 GeV.

\section{Conclusions}

\hspace{0.7cm}In this paper we have  studied
the helicity amplitudes and observables for the process
$\gamma \gamma \to ZZ$. Combining this with previous
 work in
\cite{gggg, gggZ, JikiaZ, gggZ-old, Jikia-g, JikiaZZ},
we   get the complete set of all  relevant formulae
for calculating the helicity amplitudes of
the  three processes $\gamma \gamma \to \gamma \gamma ~,~
\gamma Z~,~ ZZ$ in SM and SUSY. \par

The striking property of these three processes in SM
 above $\sim 300 GeV$, is that they are strongly dominated by just
the two helicity-conserving  amplitudes
$F_{\pm\pm\pm\pm}(\s,\t,\u)$ and
$F_{\pm\mp\pm\mp}(\s,\t,\u)= F_{\pm\mp\mp\pm}(\s,\u,\t)$;
which moreover are largely imaginary.
This simple structure is solely generated by the $W$-loop, which
at these energies, has  exactly the same structure as the
one expected from a  Pomeron
contribution. But the magnitude of this "weakly interacting"
$W$-loop contribution
is  much larger than any reasonable expectation
we might have for the "strongly interacting" Pomeron. If the
$LC_{\gamma \gamma}$ Collider is ever realized,
it will be amusing to check  this! \par

Furthermore, the aforementioned simple properties of the SM amplitudes
of  the above processes,  should make them a very  efficient tool in
searching for New Physics (NP) involving substantial imaginary
amplitudes. As a first example here and in \cite{gggg, gggZ}
we studied the contributions from loops involving charginos or
sleptons or the stop squark, in SUSY models with no
CP violating phases beyond the SM ones.
Thus, these first examples have been only applied
to energies above the threshold for  sparticle production.\par

Such measurements should be particularly useful when
we would confront the
question of identifying the nature of any possible SUSY
candidate. If such a stage is ever reached, then these loop
effects, being less (or at least differently depending) on the
soft SUSY breaking parameters, would supply important information
on the nature of such candidates. Particularly clear is the
distinction between a gaugino-type chargino which should give an
observable effect to all the three processes above; as opposed to
$\tilde t_1$ contribution which should only be visible at
$\gamma \gamma \to ZZ$; provided of course that these SUSY
particles are not too heavy. Similarly, a higgsino type chargino
with mass arround 100 GeV, will only be  visible at
$\gamma \gamma \to \gamma \gamma $ \cite{gggg}. \par

The standard SUSY scenarios we have explored in the present
and previous papers \cite{GPRgamma, gggg, gggZ}, certainly
do not exhaust  the possibilities to use
$\gamma \gamma \to \gamma \gamma ~,~\gamma Z ~,~ZZ$,  in order to
probe new physics.
They should certainly exist many more, particularly   related
to complex phases, that the NP amplitudes might for some
reason have \cite{GPRgamma}. Within the SUSY framework, the next
thing of this type that comes to mind, is to explore
the sensitivity to the  CP violating phases
affecting the soft SUSY breaking parameters.
This should affect both chargino and stop contributions.
Furthermore, in explorations of the SUSY parameter space
away from the decoupling limit, contributions
from the heavier  CP-even $H^0$-pole may also affect
$\gamma \gamma \to ZZ$,  providing useful information.\par

The overwhelming dominance of the imaginary parts of the
two helicitity conserving
amplitudes in $\gamma \gamma \to \gamma \gamma ~,~\gamma Z ~,~ZZ$
at high energies in SM, is simply so strikingly exclusive,
that it cannot stand without some  useful consequences. This
constitutes a strong motivation for the achievement of high
energy polarized photon-photon collisions.\par

\newpage

\renewcommand{\theequation}{A.\arabic{equation}}
\renewcommand{\thesection}{A.\arabic{section}}
\setcounter{equation}{0}
\setcounter{section}{0}

{\large \bf Appendix A: The $\gamma \gamma \to Z Z $
helicity amplitudes in the Standard and SUSY models.}

The invariant helicity amplitudes for  the process
\bq
\gamma (p_1,\lambda_1) \gamma (p_2,\lambda_2) \to
Z (p_3,\lambda_3) Z (p_4,\lambda_4) \ \ ,
\label{ggZZ-process}
\eq
are denoted as\footnote{Their sign is related to the sign of
the $S$-matrix through   $S_{\lambda_1 \lambda_2
\lambda_3\lambda_4}= 1+i (2\pi)^4 \delta(p_f-p_i)
F_{\lambda_1 \lambda_2 \lambda_3\lambda_4}$. }
$F_{\lambda_1 \lambda_2 \lambda_3\lambda_4}(\beta_Z,\t,\u)$,
where the momenta and
helicities of the incoming  photons and outgoing $Z$'s
 are indicated in parentheses, and the definitions
\bq
\s=(p_1+p_2)^2 = \frac{4\mzd}{1-\beta_Z^2}
~~ ~,~ ~ \t=(p_1-p_3)^2 ~ ~,~ ~\u=(p_1-p_4)^2 ~ ,
\label{kin1}
\eq
\bq
\s_4=\s-4\mzd ~, ~ \s_2=\s-2\mzd ~, ~ \t_1=\t-\mzd
~,~ \u_1=\u- \mzd ~  ~ \label{kin2}
\eq
are used. The parameter   $\beta_Z$ in (\ref{kin1}) coincides with
the $Z$-velocity in the $ZZ$ c.m. frame, and it is convenient to
be used instead of  $\s$. Denoting by $\vartheta^*$
the c.m. scattering angle of $\gamma \gamma \to ZZ$,
we also note
\bqa
\t=\mzd -\frac{\s}{2}(1-\beta_Z \cos\vartheta^*) & , &
\u=\mzd -\frac{\s}{2}(1+\beta_Z \cos\vartheta^*) ~~ ,
\label{kin3} \\
Y=\t \u -\mz^4=\frac{s^2\beta_Z^2}{4}  \sin^2 \vartheta^*=\s p_{TZ}^2
& , & \Delta =\sqrt{\frac{\s \mzd}{2Y}} ~ ,\label{kin4}
\eqa
where $p_{TZ}$ is the $Z$ transverse momentum.

Bose statistics, combined with the Jacob-Wick (JW)
phase conventions\footnote{This convention is not used in
\cite{JikiaZZ, GloverZZ}.} for the helicity wavefunction of the so called
second particle, demands
\bqa
F_{\lambda_1 \lambda_2 \lambda_3\lambda_4}(\beta_Z,\t,\u) &=&
F_{\lambda_2 \lambda_1 \lambda_4\lambda_3}(\beta_Z,\t,\u)
(-1)^{\lambda_3-\lambda_4} \ , \label{Bose2} \\
F_{\lambda_1 \lambda_2 \lambda_3\lambda_4}(\beta_Z,\t,\u) &=&
F_{\lambda_2 \lambda_1 \lambda_3\lambda_4}(\beta_Z,\u,\t)
(-1)^{\lambda_3-\lambda_4} \ , \label{Bose1} \\
F_{\lambda_1 \lambda_2 \lambda_3\lambda_4}(\beta_Z,\t,\u) &=&
F_{\lambda_1 \lambda_2 \lambda_4\lambda_3}(\beta_Z,\u,\t)
 \ ; \label{Bose12}
\eqa
while the standard form of the $Z$ polarization vectors implies
the constraint
\bq
F_{\lambda_1 \lambda_2 \lambda_3\lambda_4}(\beta_Z,\t,\u) =
F_{\lambda_1 \lambda_2, -\lambda_3, -\lambda_4}(-\beta_Z,\t,\u)
(-1)^{\lambda_3-\lambda_4} \ . \label{-beta}
\eq
Finally, parity invariance  implies
\bq
F_{\lambda_1 \lambda_2 \lambda_3\lambda_4}(\beta_Z ,\t,\u) =
F_{-\lambda_1,-\lambda_2,- \lambda_3,-\lambda_4}(\beta_Z,\t,\u)
(-1)^{\lambda_3-\lambda_4} \  .  \label{parity}
\eq

As a result, the 36  helicity amplitudes
may be expressed in terms of just the eight independent ones
\bqa
& F_{+++-}(\beta_Z,\t,\u)~ , ~ F_{++++}(\beta_Z,\t,\u), &
  F_{+-++}(\beta_Z,\t,\u)~, ~ F_{+-00}(\beta_Z,\t,\u)~, \nonumber \\
&F_{++00}(\beta_Z,\t,\u)~ ,~  F_{+++0}(\beta_Z, \t,\u), &
F_{+-+0}(\beta_Z,\t,\u)~ , ~  F_{+-+-}(\beta_Z,\t,\u )~.
\label{8basic}
\eqa
Using these  and (\ref{-beta}), we determine
\bqa
F_{++--}(\beta_Z, \t,\u) & = & F_{++++}(-\beta_Z, \t, \u)
\  \ , \label{-beta++--} \\
F_{++-0}(\beta_Z, \t,\u) & = & F_{+++0}(-\beta_Z, \t, \u)
\  \ , \label{-beta++-0}
\eqa
while the remaining 26 amplitudes may be obtained from the
ten in (\ref{8basic}, \ref{-beta++--}, \ref{-beta++-0}),
by $(\t \leftrightarrow \u)$ interchanges
or helicity changes; compare
(\ref{Bose2}-\ref{Bose12}, \ref{parity}). \par

In SM or any SUSY model, there are two types of contributions
to these amplitudes. The first type consists of
the one-loop  diagrams involving four external legs,
like those  contributing to the
$\gamma \gamma \to \gamma \gamma$ and
$\gamma \gamma \to \gamma Z$ processes
\cite{gggZ, JikiaZ, gggg, GPRgamma}; while the second type
includes the Higgs s-channel pole contributions, arising from
loops   with three external legs
generating\footnote{Here $h^0$ denotes
any  neutral Higgs boson.}  $h^0 \gamma \gamma$
interactions   \cite{JikiaZZ}.
To express them economically, we use the notation of \cite{Hagiwara}
for the $B_0$, $C_0$ and $D_0$
1-loop functions first defined by Passarino and Veltman
\cite{Passarino}. For brevity, we introduce the shorthand
writing\footnote{The numbers used in the notation of the
one loop functions, correspond to the momenta of
process (\ref{ggZZ-process}), (taken here as incoming).}

\bqa
B_0(\s)& \equiv  & B_0(\s;m,m) \ , \label{B0} \\
C_0(\s) & \equiv & C_0(12)=C_0(0,0,\s;m,m,m) \ , \label{C0}
\eqa
\bqa
B_Z(\s)& \equiv  & B_0(\s)-B_0(m^2_Z +i\epsilon) \ , \label{BZ} \\
C_Z(\t) & \equiv & C_0(13)\equiv C_0(24) \equiv C_0(0,\mzd,\t;m,m,m)
\ , \label{CZ} \\
C_{ZZ}(\s) &\equiv & C_0(34)\equiv C_0(\mzd, \mzd, \s; m,m,m)
\ , \label{CZZ} \\
D_{ZZ}(\s,\u) & \equiv & D_0(123)\equiv
D_0(0,0,\mzd,\mzd,\s,\u;m,m,m,m)= \nonumber \\
D_{ZZ}(\u,\s) &\equiv  & D_0(321)\equiv
D_0(\mzd, 0,0,\mzd,\u,\s;m,m,m,m) \  \label{DZZsu} , \\
D_{ZZ}(\s,\t) & \equiv & D_0(213)\equiv
D_0(0,0,\mzd,\mzd,\s,\t;m,m,m,m)= \nonumber \\
D_{ZZ}(\t,\s) &\equiv  & D_0(312)\equiv
D_0(\mzd, 0,0,\mzd,\t,\s;m,m,m,m) \  \label{DZZst} , \\
D_{ZZ}(\t,\u) & \equiv & D_0(132)\equiv
D_0(0,\mzd,0, \mzd,\t,\u;m,m,m,m)= \nonumber \\
D_{ZZ}(\u,\t) &\equiv  & D_0(231)\equiv
D_0(0, \mzd, 0,\mzd,\u,\t;m,m,m,m) \  \label{DZZtu}
\eqa
In diagrams with four external legs, the expressions
\bqa
\tilde F(\s,\t,\u)& \equiv  & D_{ZZ}(\s,\t)+D_{ZZ}(\s,\u)+D_{ZZ}(\t,\u)
\ \ , \label{Ftilde} \\
E_1(\s,\t) & \equiv  & 2\t_1 C_Z(\t) -\s\t D_{ZZ}(\s,\t) ,
\label{E1} \\
E_2(\t,\u)&\equiv & 2\t_1 C_Z(\t)+2 \u_1 C_Z(\u) - Y D_{ZZ}(\t,\u) ,
\label{E2}
\eqa
often appear in the amplitudes below. \par

\vspace{1.0cm}

The \underline{ neutral Higgs-pole} contribution to the
$\gamma \gamma \to ZZ$ helicity amplitudes,  involve
the $h^0\gamma \gamma$  interaction  generated by spin-1,
spin-1/2 or spin-0 loops. They are concisely described  as
\cite{Gunion-book}
\bqa
&& F^h_{\lambda_1 \lambda_2 \lambda_3\lambda_4}(\gamma \gamma \to ZZ)
= -~\frac{\alpha^2}{2\swd\cwd}\left \{ \sum_i \H_i(\tau)
\right \}\frac{\s}{\s-\mh^2+i \mh \Gamma_h} \cdot \nonumber \\
&& \cdot \frac{(1+\lambda_1 \lambda_2)}{2}
~ \left [(1+\lambda_3 \lambda_4)\, \frac{\lambda_3 \lambda_4}{2}~ -
~\frac{1+\beta_Z^2}{1-\beta_Z^2}~ (1- \lambda_3^2)(1- \lambda_4^2)
\right ] \ , \label{Fh}
\eqa
where  the index $i$ runs over the
particles in the loop describing the $h^0\gamma \gamma$
vertex, whose spin is   (1, 1/2 or 0). In (\ref{Fh})
the $h^0ZZ$ coupling is taken as in SM; which means \eg\@ that an extra
factor $\sin(\beta-\alpha)$ should be introduced in the case
of  the lightest CP even  SUSY Higgs particle.
If  the interaction Lagrangian
of the neutral Higgs   to a charged particle pair with spin
(1, 1/2, 0) is given by   \cite{Gunion-book}
\bq
\L_{int} = -~\frac{g m_f}{2 \mw} \bar \psi \psi h^0
~+ ~ g \mw W^+_\mu W^{\mu-} h^0 ~-~ \frac{g m_{H^\pm}^2}{\mw} H^+H^-
h^0 \ \ , \label{h-int}
\eq
then
\bq
\H_i(\tau) =N_{ci}Q^2_iF_i(\tau) ~ , ~ \label{Hitau}
\eq
 with
\bqa
F_1(\tau) & =& \frac{2\mh^2}{\s} ~+ 3 \tau + 3 \tau
\left ( \frac{8}{3} ~-~\frac{2 \mh^2}{3\s} ~-\tau \right) f(\tau)
\ ,  \label{F1} \\
F_{1/2}(\tau) &=& - 2\tau [1+ (1-\tau)f(\tau)] \ , \label{F12}\\
F_0(\tau) & =& \tau [1- \tau f(\tau)] \ , \label{F0}
\eqa
where (compare (\ref{C0}))
\bq
\tau= ~\frac{4 m_i^2}{\s} ~~~~, ~~~~ f(\tau)=~-\frac{\s}{2} C_0(\s)
\ . \label{ftau}
\eq
In (\ref{Hitau}),  $Q_i$ is the charge  and $N_{ci}$ the colour
multiplicity of the particle contributing to
$h^0\gamma \gamma $. If more than one neutral Higgs
particle
with couplings of the type given in (\ref{h-int}) exists,
then a summation over their contributions
should be included in  (\ref{Fh}).\par

\vspace{1.0cm}

We next turn to the contribution from loops in diagrams
involving four external legs. It is easiest to describe them  by using a
non-linear gauge as in \cite{Dicus}, for which the
same type of particle propagates along the entire
loop\footnote{For this gauge, the couplings $\gamma W^\pm\phi^\mp$,
$Z W^\pm\phi^\mp$ vanish.}.
Thus, the various contributions may simply be described as
arising from loops due to a scalar particle, a $W$ boson or
a fermion. We give them in this order below.

\vspace{0.5cm}
\underline{The scalar} particle loop contribution to the helicity
amplitudes. We consider the loop contribution
due to a scalar particle of mass $m$,
charge $Q_{S}$ and a definite value of third isospin component
$t^S_3$.  In analogy to (A.36) of  \cite{gggZ}, this
contribution is  written as
\bq
F^S_{\lambda_1 \lambda_2 \lambda_3\lambda_4}(\beta_Z,\t,\u)\equiv
\alpha^2 Q^2_S \left (g^Z_{S}\right )^2
A^S_{\lambda_1 \lambda_2 \lambda_3\lambda_4} (\beta_Z,\t,\u; m) \ ,
\label{FSamp}
\eq
where
\bq
g^Z_{S}  =  \frac{t^S_3-Q_S \sw^2}{\sw\cw}  \ \ . \label{gSZ}
\eq
Relations (\ref{FSamp} , \ref{gSZ}) are  directly applicable
to a purely L-  or R-slepton or squark,
while the appropriate mixing should
be taken into account in a case like a stop contribution.
The scalar contributions to the r.h.s. of (\ref{FSamp}) for the eight
basic amplitudes in (\ref{8basic}),  are:
\bqa
&&A^S_{+++-}(\beta_Z, \t, \u ; m)=
-\, \frac{4 \s_2 Y}{\t_1\u_1\s_4}
+\, \frac{4 \s_2 m^2 (\s\s_4 -2 Y)}{\s_4 Y} C_0(\s)
+\, \frac{4 \s\s_4 m^2}{Y} C_{ZZ}(\s)
 \nonumber \\
&&+~ 8 m^4 \Fstu +\frac{4 [\mzd Y- m^2 \s\s_4]}{\s^2\s_4}
\Etu -
\frac{8 \mzd m^2 Y}{\s\s_4} D_{ZZ}(\t,\u)
\nonumber \\
&& -~4 \Bigg \{ \frac{m^2 \t}{Y}  \Est +
2 m^2 \left (1+\, \frac{\mzd \s_2}{\s_4 \t_1} \right ) C_Z(\t)
+\, \frac{\mzd Y}{\s_4 \t_1^2}
\left ( \frac{2\t}{\s}-1 \right ) B_Z(\t)
 \nonumber \\
&& +~ \frac{2 \mz^4 m^2}{\s_4} D_{ZZ}(\s, \t)
~~ + ~~(\t \leftrightarrow \u) \Bigg \} \ , \label{S+++-}
\eqa
\bqa
&& A^S_{++++}(\beta_Z, \t, \u ; m)=
\frac{4 [\mzd (2Y-\s \s_4)+\beta_Z \s Y]}{\s_4 \t_1 \u_1}
~+~\frac{16 \mzd m^2}{\s_4}\, C_0(\s) + 8m^4 \tilde F(\s,\t,\u)
   \nonumber \\
&& + \frac{8 Y m^2}{\s \s_4} (\s_2 + \beta_Z \s )D_{ZZ}(\t, \u) -
\frac{2 [ (\s_2  +\beta_Z \s) Y -4 \s \mzd m^2]}{\s^2 \s_4} E_2(\t,\u)
  \nonumber \\
&& + 4 \Bigg \{\frac{2 m^2 \mz^4}{\s_4} D_{ZZ}(\s,\t) -
\frac{[\s_2+\beta_Z \s][2\mzd Y +
\t_1(2 \t_1 +\s) (\t+\mzd)]}{2\s_4 \s \t_1^2} B_Z(\t)
\nonumber \\
&& -~ \frac{2 m^2[\t_1 (\t-\u)+Y](\s_2+\beta_Z\s)}{\s_4 \t_1
\s}C_Z(\t) ~~+ ~~(\t \leftrightarrow \u) \Bigg \}\ , \label{S++++}
\eqa
\bqa
&& A^S_{+-++}(\beta_Z, \t, \u ; m)=
-~\frac{4\s_2 Y}{\s_4 \t_1 \u_1}~+~
\frac{4 (\s_4 m^2 +\mz^4)}{\s_4 Y}
\left [ \s\s_2 C_0(\s) +(\s\s_4 -2Y) C_{ZZ}(\s) \right ]
\nonumber \\
&& -~ \frac{4 m^2 \s_2}{\s\s_4}\Etu +8m^4 \Fstu
+ 4 \Bigg \{ \frac{\mzd (Y+2\t \mzd)}{\s_4 \t_1^2} B_Z(\t)
 -\, \frac{2m^2 \s_2 \t}{\s_4 \t_1}C_Z(\t)
\nonumber \\
&& +~\frac{2m^2 \mz^4}{\s_4} D_{ZZ}(\s,\t)
 -\,\frac{\t(\s_4 m^2 +\mz^4)}{\s_4 Y} \Est
 ~~+ ~~(\t \leftrightarrow \u) \Bigg \}\ , \label{S+-++}
\eqa
\bqa
&& A^S_{+-00}(\beta_Z, \t, \u ; m)=
-~\frac{16 \mzd Y}{\s_4 \t_1 \u_1}
+~ \frac{2\s^2 \s_2 \mzd}{\s_4 Y} C_0(\s)
+~ \frac{2\s \mzd}{\s_4 Y}(\s \s_4 -2Y) C_{ZZ}(\s)
\nonumber \\
&& -~\frac{4 (\t-\u)^2\mzd m^2}{\s\s_4} D_{ZZ}(\t,\u)
-4 \Bigg \{ \frac{2\mzd}{\s_4 \t_1^2} (\t^2 +\mz^4) B_Z(\t)
-~\frac{8 m^2 \mzd Y}{\s_4 \s \t_1} C_Z(\t)
\nonumber \\
&& -~ \frac{\s\mzd m^2}{\s_4} D_{ZZ}(\s,\t)
+~\frac{\s\t\mzd}{2 \s_4 Y} \Est
~~+ ~~(\t \leftrightarrow \u) \Bigg \}\ , \label{S+-00}
\eqa
\bqa
&& A^S_{++00}(\beta_Z, \t, \u ; m)=
-~\frac{4 \mzd m^2 (\t-\u)^2}{\s\s_4} D_{ZZ}(\t,\u)
+~\frac{16 \mzd Y}{\s_4 \t_1 \u_1}
+~\frac{32 \mzd m^2}{\s_4} C_0(\s)
\nonumber \\
&& +~\frac{2\mzd (\t-\u)^2}{\s^2 \s_4} \Etu
- 4 \Bigg \{ \frac{2 \mzd}{\s_4 \s \t_1^2}
[2 \mzd Y +\t_1 (\t-\u)(\t+\mzd)]B_Z(\t)
\nonumber \\
&& -~\frac{8 \mz^4 m^2}{\s_4 \t_1} C_Z(\t)
-~ \frac{\s \mzd m^2}{\s_4} D_{ZZ}(\s,\t)
~~+ ~~(\t \leftrightarrow \u) \Bigg \}\ , \label{S++00}
\eqa
\bqa
&& A^S_{+++0}(\beta_Z, \t, \u ; m)/\Delta=
- 4~\frac{(\t-\u)}{\s_4} \Bigg \{
\frac{(1+\beta_Z)Y}{\t_1\u_1}~ + 2m^2 C_0(\s)
\nonumber \\
&& -~\frac{1}{\s}
\left [ \frac{Y(1+\beta_Z)}{2\s} +\beta_Z m^2 \right ] \Etu
+~\frac{(1+\beta_Z) m^2 Y}{\s} D_{ZZ}(\t,\u) \Bigg \}
\nonumber \\
&& +~\frac{4}{\s_4} \Bigg \{
\frac{(1+\beta_Z) Y}{\s t_1^2}(\s \mzd -2 \t \t_1)B_Z(\t)
+~\frac{2 m^2 (1+\beta_Z)(\t^2 -\mz^4+Y)}{\t_1}C_Z(\t)
\nonumber \\
&& + m^2 (Y+\t^2-\mz^4)D_{ZZ}(\s,\t)
~~- ~~(\t \leftrightarrow \u) \Bigg \}\ , \label{S+++0}
\eqa
\bqa
&& A^S_{+-+0}(\beta_Z, \t, \u ; m)/\Delta=
-~\frac{4 (\u-\t -\s \beta_Z)Y}{\s_4 \t_1\u_1}
 +~ \frac{4 (\u-\t + \s \beta_Z)}{\s_4} B_Z(\s)
\nonumber \\
&& +~\frac{2\s}{\s_4 Y} \Big \{
(\t-\u) (2 \mz^4 -\s_2^2) +\beta_Z
\left [ 4 m^2 Y +\s (\t^2 +\u^2) \right ] \Big \} C_0(\s)
\nonumber \\
&& +~ \frac{2 \s\s_2}{\s_4 Y}
\left \{ (\u-\t)\s_4 +\beta_Z (\s\s_4 -2Y)\right \} C_{ZZ}(\s)
+~\frac{4m^2 (\t-\u)}{\s\s_4}\Etu
\nonumber \\
&& -4 \Bigg \{\frac{\mzd Y -\t\t_1(\t+\mzd)+\beta_Z(\mzd Y -\t\t_1^2)}
{\s_4 \t_1^2} B_Z(\t)
\nonumber \\
&& -~\frac{[(2 \mz^4+\t\s_2)(2m^2 Y +\s\t^2)+
\beta_z\s\t (4m^2 Y +\s\t^2)]}{2\s_4 Y\s\t} \Est
\nonumber \\
&& +~\frac{2m^2[(2 \mzd \t_1 +\s\t)Y -\beta_Z \s^2\t^2]}
{\s\s_4\t\t_1}  C_Z(\t)
~~- ~~(\t \leftrightarrow \u ~, ~ \beta_Z \to -\beta_Z)
\Bigg \}\ , \label{S+-+0}
\eqa
\bqa
&& A^S_{+-+-}(\beta_Z, \t, \u ; m)=
\frac{4[\s_2 Y+\beta_Z \mzd \s (\u-\t)]}{\s_4 \t_1 \u_1}
-\, \frac{4\s_2 [Y-\s \{\s_4+\beta_Z (\u-\t)\}]}{\s_4 Y}B_Z(\s)
\nonumber \\
&& +\, \frac{4\s\s_2}{\s_4 Y}
\Bigg \{\left [m^2+~\frac{\s(\s\s_4 -Y+\mz^4)}{2 Y} \right ]
[\s_4 +\beta_Z (\u-\t)] -\s\s_4 -\mz^4 \Bigg \} C_0(\s)
\nonumber \\
&& +\, \frac{4\s}{Y} \Bigg \{
\Big [ m^2 +\,
\frac{\t^2 (\t^2-\mz^4 +Y) +\u^2 (\u^2-\mz^4 +Y)+2
Y(\s_2^2-\mz^4)}{2Y\s_4} \Big ] [\s_4 +\beta_Z(\u-\t) ]
\nonumber \\
&& +\mz^4 -\s_2^2 \Bigg \} C_{ZZ}(\s)
+ 8 m^2 \Big (m^2 -\, \frac{\mzd Y}{\s\s_4} \Big )D_{ZZ}(\t,\u)
\nonumber \\
&& +4 \Bigg \{
-\, \frac{\t}{2 \s_4 Y^2} [2 m^2 \s_4 Y +\s\s_4\t^2 - 2\mz^4 Y
-\beta_Z \s\t(\t^2 -\mz^4 +Y)]\Est
\nonumber \\
&& +\, \frac{m^2}{\s_4 Y}\, [2 m^2 \s_4 Y+ \s \s_4 \t^2
- 2\mz^4 Y-2\beta_Z\s\t (\t^2-\mz^4+Y)] D_{ZZ}(\s,\t)
\nonumber \\
&& +\,\frac{2m^2}{\s_4\t_1} \Big [
\mzd (\s_4+\beta_Z\s) -\,\frac{2 \mzd Y}{\s}\,-\,
\frac{\beta_Z\s\t (\t^2-\mz^4)}{Y} \Big ] C_Z(\t)
\nonumber \\
&& + \Big [ \Big (\frac{\mz^4 (\t-\u)}{\s_4\t_1^2}-\frac{1}{2} \Big )
(1+\beta_Z)  +\, \frac{\mzd}{\s_4} \Big (1-\frac{2\mz^4}{\t_1^2} \Big )
+\, \frac{2\t \beta_Z}{\s_4}
\nonumber \\
&&  -\, \frac{\t^2}{Y\s_4}
\bigg (\s_4 -\beta_Z (\t-\u) \bigg )\Big ]B_Z(\t)
~~+ ~~(\t \leftrightarrow \u ~, ~ \beta_Z \to -\beta_Z)
\Bigg \}\ . \label{S+-+-}
\eqa

\vspace{0.5cm}
\underline{The $W$}  loop contribution to the helicity amplitudes
are generated in the non-linear gauge \cite{Dicus}, by  loops
involving $W$, Goldstone bosons and FP ghosts, in diagrams involving
four external legs. They  have first been presented
in \cite{JikiaZZ}, and have also been calculated in
\cite{ZZloop}. Here we give a new expression, using
the results in (\ref{S+++-}-\ref{S+-+-}). The $W$-loop
contribution to the $\gamma \gamma \to ZZ$ helicity amplitudes is
thus written as
\bq
F^W_{\lambda_1 \lambda_2 \lambda_3\lambda_4}(\beta_Z,\t,\u)\equiv
\frac{\alpha^2}{\swd}
A^W_{\lambda_1 \lambda_2 \lambda_3\lambda_4} (\beta_Z,\t,\u)
\label{FWamp}
\eq
with
\bqa
A^W_{\lambda_1 \lambda_2 \lambda_3\lambda_4} (\beta_Z,\t,\u)
&= &
\frac{(12\cw^4-4\cwd +1)}{4\cwd}
A^S_{\lambda_1 \lambda_2 \lambda_3\lambda_4} (\beta_Z,\t,\u; \mw)
\nonumber \\
&+& \delta^W_{\lambda_1 \lambda_2 \lambda_3\lambda_4} (\beta_Z,\t,\u)
\ , \label{deltaW}
\eqa
and
\bq
\delta^W_{+++-} (\beta_Z,\t,\u)=0 \ , \label{W+++-}
\eq
\bqa
\delta^W_{++++} (\beta_Z,\t,\u)& = &
 \frac{8 \mzd \s \beta_Z}{\s_4} C_0(\s)+\,
\frac{4 [2 \cwd \s_4 (\s_2 +\beta_Z \s)+
\mzd (\s_4+\beta_z \s)  ]}{\s\s_4} \Etu
\nonumber \\
& -&  4 \cwd [2 \mz^4 +(4\mwd -\s)(\s_2+\beta_Z \s)]\Fstu
\nonumber \\
&+ &\frac{2\s\mzd}{\s_4} (\s_4+\beta_Z\s_2)
[D_{ZZ}(\s,\t)+D_{ZZ}(\s,\u)] ~ , \label{W++++}
\eqa
\bqa
&& \delta^W_{+-++} (\beta_Z,\t,\u) =
\frac{4\s}{\s_4 Y}[\s_2\s_4 (4\mwd -\mzd)-8 Y \mwd]C_0(\s)
\nonumber \\
&&+\, \frac{4 (4\mwd -\mzd)(\s\s_4 -2Y)}{Y} C_{ZZ}(\s)
+8 \mwd\Big [ 4\mwd-\mzd +\, \frac{2 Y}{\s_4}\Big ]\Fstu
\nonumber \\
&& + \Bigg \{
\frac{4 [4 \mwd (\t+\mzd)^2+\t\s_4\mzd]}{\s_4 Y} \Est
+ \, \frac{4\mzd Y}{\s_4} D_{ZZ}(\s,\t)
~~+ ~~(\t \leftrightarrow \u) \Bigg \}\ , \label{W+-++}
\eqa
\bqa
&& \delta^W_{+-00}(\beta_Z,\t,\u) =
\frac{2\s}{\s_4 Y} [8\mwd (\s_2\s_4 -4Y) +\s_4 (\s_2^2
-2Y)]C_0(\s)
\nonumber \\
&&  +\, \frac{2}{Y}(\s_2+8\mwd) (\s \s_4- 2Y)C_{ZZ}(\s)
 + 4\mwd (\s_2+8 \mwd)\Fstu
 -\, \frac{2}{\s_4}(\s_4+16 \mwd)\Etu
\nonumber \\
&&
+\Bigg \{
\frac{4\t_1}{\s_4 Y} \Big [8 \mwd \Big (Y +(\t+\mzd)^2 \Big )
-\s_2\s_4 \t \Big ] C_Z(\t) +\, \frac{2}{\s_4 Y}
\Big [ 8 \mwd (\t^2\s\s_4 -2 Y \t_1^2)
\nonumber \\
&&  +\s
\Big ( (\t^2-\mz^4)^2 -2 \mzd \t^2 \s_4 -\s\t Y \Big )\Big ]
D_{ZZ}(\s,\t)
~~+ ~~(\t \leftrightarrow \u) \Bigg \}\ , \label{W+-00}
\eqa
\bqa
 \delta^W_{++00}(\beta_Z,\t,\u)& = &
4\s C_0(\s) -4 \mwd (\s+ 2\mzd -8\mwd )\Fstu
\nonumber \\
&- & \frac{4(4\mwd -\mzd)}{\s}\Etu \ ,\label{W++00}
\eqa
\bqa
 \delta^W_{+++0}(\beta_Z,\t,\u)/\Delta & = &
- \, \frac{(\t-\u)(\s_4+\beta_Z \s)}{\s\s_4}
\Big [2\s C_0(\s) +\Etu \Big ]
\nonumber \\
&+& \frac{(\s_4 +\beta_Z \s)}{\s_4}
\Bigg \{ (\t^2 -\mz^4 +Y)D_{ZZ}(\s,\t)
~~- ~~(\t \leftrightarrow \u) \Bigg \}\ , \label{W+++0}
\eqa
\bqa
&& \delta^W_{+-+0}(\beta_Z,\t,\u)/\Delta  =
-\, \frac{2}{\s_4}
\Big [ (\u-\t-\beta_Z \s)\s_4 +8 \cwd \s (\u-\t+\beta_Z\s_4)
\Big ] C_0(\s)
\nonumber \\
&& +16 \cwd (\t-\u-\beta_Z \s)C_{ZZ}(\s)- \,
\frac{64 \cwd \mwd \beta_Z Y}{\s_4}\Fstu
+\, \frac{8\cwd}{\s_4} Y(\u-\t-\beta_Z\s)D_{ZZ}(\t,\u)
\nonumber \\
&& +\, \frac{(\t-\u+\beta_Z\s)}{\s}\Etu
-\, \Bigg \{
\frac{(\u-\t-\beta_Z \s)}{\s_4}
\Big [ 2\mz^4 +2\t^2 +\s\t -8 \cwd Y \Big ] D_{ZZ}(\s,\t)
\nonumber \\
&& -\, \frac{16\cwd}{\s_4}(\t+\mzd)(1+\beta_Z) \Est
~~- ~~(\t \leftrightarrow \u ~, ~ \beta_Z \to -\beta_Z)
\Bigg \}\ , \label{W+-+0}
\eqa
\bqa
&& \delta^W_{+-+-}(\beta_Z,\t,\u)  =
16\cwd \s \Big [\frac{\s_2}{\s_4} C_0(\s)+C_{ZZ}(\s) \Big ]
\nonumber \\
&& +\, \frac{4 \cwd}{\s_4} \Big [ \s(\s_2+4\mwd)(\s_4
+\beta_Z[\t-\u]) -2Y\s_2 \Big ]\Fstu
+\Bigg \{ \frac{8 \cwd (\s_2+\beta_Z \s)}{\s_4} \Est
\nonumber \\
&& +\, \frac{2 \mzd}{\s_4} \Big [\s(\s_4+\beta_Z[\t-\u])
-2 Y \Big ]D_{ZZ}(\s,\t)
~~+ ~~(\t \leftrightarrow \u ~, ~ \beta_Z \to -\beta_Z)
\Bigg \}\ , \label{W+-+-}
\eqa

We have checked that the above $W$ loop contributions to the helicity
amplitudes agree with those of \cite{JikiaZZ}, except for a minor
misprint in the $A^W_{+-+0}$ case\footnote{We find that the term
$-24 \cwd \mwd u_1$ in the coefficient
of $C(t)$ in (3.14) of \cite{JikiaZZ}, should be
replaced by $-24 \cwd \mwd u_1/(s s_4)$.}.
It should be noticed also  that our definitions of $\t$ and $\u$
should be interchanged when comparing with \cite{JikiaZZ}, and
that these authors do not use the JW convention. \par

\vspace{0.5cm}
\underline{The fermion} loop contribution.
If the effective $Zf\bar f$ interaction is written as
\bq
\L_{Zff} = -e Z^\mu \bar f
(\gamma_\mu g_{Vf}^Z- \gamma_\mu \gamma_5 g_{Af}^Z) f ,
\label{LZff}
\eq
then the fermion loop contribution
(for a fermion of mass $m_f$),
to the $\gamma \gamma \to ZZ$ helicity amplitude,
is given by\footnote{As far as the sign of these amplitudes,
we agree with \cite{JikiaZZ}, apart from the trivial changes
introduced by our using of the JW  phase conventions.}
    \cite{GloverZZ}
\bq
F^f_{\lambda_1 \lambda_2 \lambda_3\lambda_4}(\beta_Z,\t,\u)\equiv
\alpha^2 Q^2_f \Bigg \{ (g^Z_{vf})^2
A^{vf}_{\lambda_1 \lambda_2 \lambda_3\lambda_4} (\beta_Z,\t,\u; m_f )
+(g^Z_{af})^2
A^{af}_{\lambda_1 \lambda_2 \lambda_3\lambda_4} (\beta_Z,\t,\u; m_f)
\Bigg \} \ .
\label{Ffamp}
\eq
In SM, the vector and axial vector couplings for the quarks and
leptons  are given by (\ref{LZff}),
\bq
g_{vf}^Z=\frac{t_3^f-2Q_f\sw^2}{2\sw\cw} ~~~~~ ,
~~~~~ g_{af}^Z=\frac{t_3^f}{2\sw\cw} ~ ~~~~ ,
\label{gvafZ}
\eq
where $t_3^f$ is the  third isospin component of the fermion,
and  $Q_f$ is its charge.

The vector and axial contributions to the fermion loop amplitudes
in (\ref{Ffamp}), may be expressed in terms of the $A^S$
amplitudes of (\ref{S+++-}-\ref{S+-+-}), by
\bqa
A^{vf}_{\lambda_1 \lambda_2 \lambda_3\lambda_4}
(\beta_Z,\t,\u; m_f)& =&  -2
A^{S}_{\lambda_1 \lambda_2 \lambda_3\lambda_4} (\beta_Z,\t,\u; m_f )
+\delta^{vf}_{\lambda_1 \lambda_2 \lambda_3\lambda_4}
(\beta_Z,\t,\u; m_f)  , \label{deltafv} \\
 A^{af}_{\lambda_1 \lambda_2 \lambda_3\lambda_4}
(\beta_Z,\t,\u; m_f)& = & -2
A^{S}_{\lambda_1 \lambda_2 \lambda_3\lambda_4} (\beta_Z,\t,\u; m_f )
+\delta^{af}_{\lambda_1 \lambda_2 \lambda_3\lambda_4}
(\beta_Z,\t,\u; m_f)  , \label{deltafa}
\eqa
where
\bq
\delta^{vf}_{+++-}(\beta_Z,\t,\u; m_f)=
\delta^{af}_{+++-}(\beta_Z,\t,\u; m_f)=0 ~ ,
\label{fva+++-}
\eq
\bq
\delta^{vf}_{++++}(\beta_Z,\t,\u; m_f)=
4(\s_2+\s \beta_Z) [m_f^2 \Fstu -\frac{1}{2\s}\Etu ]
~ , \label{fv++++}
\eq
\bqa
&& \delta^{af}_{++++}(\beta_Z,\t,\u; m_f)=
-\, \frac {8\s m_f^2}{\s_4}
\Big [ 4 \beta_Z C_0(\s) +
(\s_4 +\beta_Z \s_2)[D_{ZZ}(\s,\t)+D_{ZZ}(\s,\u)]\Big ]
\nonumber \\
&& + 4 \mf^2 (\s_2+\s \beta_Z +8 \mf^2)\Fstu -
\frac{2}{\s}\Big [ \s_2 +\s \beta_Z+
\frac{8\mf^2 (\s_4 +\s \beta_Z)}{\s_4}\Big ] \Etu
~ , \label{fa++++}
\eqa
\bqa
&& \delta^{vf}_{+-++}(\beta_Z,\t,\u; m_f)=
-\, \frac{4\mzd}{Y}\Big [\frac{\s(\s_2\s_4-2Y)}{\s_4}C_0(\s)+
(\s\s_4-2 Y)C_{ZZ}(\s) \Big ]
\nonumber \\
&& -\, \frac{4 \mzd}{\s_4 Y}\Big [ (\t+\mzd)^2 \Est
+(\u+\mzd)^2 \Esu \Big ]-
8 \mzd \mf^2 \Fstu ~, \label{fv+-++}
\eqa
\bqa
&& \delta^{af}_{+-++}(\beta_Z,\t,\u; m_f)=
8\s \Big [ \frac{\s_2 (4\mf^2 -\mzd)}{2Y}+\frac{\mzd}{\s_4}
\Big ]C_0(\s) +\frac{4(\s\s_4-2 Y)}{Y}(4 \mf^2-\mzd)C_{ZZ}(\s)
\nonumber \\
&& + 8\mf^2 (4\mf^2 -\mzd)\Fstu +\frac{16 \mf^2
Y}{\s_4}D_{ZZ}(\t,\u)
\nonumber \\
&& -\, \frac{4}{Y}
\Bigg \{ \Big [\frac{\mzd(\t+\mzd)^2}{\s_4}+4 \mf^2 \t \Big ] \Est
+  \Big [\frac{\mzd(\u+\mzd)^2}{\s_4}+4 \mf^2 \u \Big ] \Esu
\Bigg \}\ , \label{fa+-++}
\eqa
\bq
\delta^{vf}_{++00}(\beta_Z,\t,\u; m_f)=
- 8 \mf^2 \mzd \Fstu  + \frac{4 \mzd}{\s} \Etu
~ , \label{fv++00}
\eq
\bqa
&& \delta^{af}_{++00}(\beta_Z,\t,\u; m_f)=
- 8 \mf^2 \Big ( \mzd +\frac{2 \s_2 \mf^2}{\mzd} \Big ) \Fstu
\nonumber \\
&& - \frac{4}{\s} (4 \mf^2 -\mzd) \Etu -
\frac{16 \mf^2 \s}{\mzd} C_0(\s)
~ , \label{fa++00}
\eqa
\bqa
 && \delta^{vf}_{+-00}(\beta_Z,\t,\u; m_f) =
-\, \frac{4\s\mzd}{\s_4 Y}(\s_2\s_4 -4 Y)C_0(\s)-
\frac{4 \mzd}{Y} (\s\s_4 -2Y) C_{ZZ}(\s)
\nonumber \\
&& -8 \mzd \mf^2 \Fstu
-  \frac{4\mzd}{\s_4 Y}
\Big [ (2\t^2 +\s\t +2 \mz^4)\Est
\nonumber \\
&& +(2\u^2 +\s\u +2 \mz^4)\Esu \Big ] ~ , \label{fv+-00}
\eqa
\bqa
 && \delta^{af}_{+-00}(\beta_Z,\t,\u; m_f) =
-\, \frac{8\mf^2}{\mzd} \Bigg \{ \s \Big (\frac{\s_2^2}{Y}-2\Big )
C_0(\s)+ \s_2 \Big (\frac{\s\s_4}{Y} -2 \Big ) C_{ZZ}(\s) -
\frac{4 \mzd Y}{\s_4}D_{ZZ}(\t,\u) \nonumber \\ && +\frac{(\t^2
+\mz^4)}{Y}\Est +\frac{(\u^2 +\mz^4)}{Y}\Esu +(2 \s_2 \mf^2
+\mz^4)\Fstu \Bigg \} \nonumber \\ && - \frac{4\mzd}{Y} \Big [
\frac{\s}{\s_4} (\s_2 \s_4 -4 Y) C_0(\s) +(\s\s_4 -2 Y) C_{ZZ}(\s)
+\frac{(2\t^2+\s\t +2\mz^4)}{\s_4}\Est
\nonumber \\
&& +\frac{(2\u^2+\s\u +2\mz^4)}{\s_4}\Esu  \Big ]
~ , \label{fa+-00}
\eqa
\bq
\delta^{vf}_{+++0}(\beta_Z,\t,\u; m_f)=0 ~, \label{fv+++0}
\eq
\bqa
&& \delta^{af}_{+++0}(\beta_Z,\t,\u; m_f)/\Delta =
- \, \frac{4 (\s_4+\beta_Z \s)\mf^2}{\s_4 \mzd}
\Bigg \{ (\u-\t) \Big [2 C_0(\s) +\frac{1}{\s} \Etu \Big ]
\nonumber \\
&& - (\s\t -2 \mzd \t_1)D_{ZZ}(\s, \t) +
(\s\u -2 \mzd \u_1)D_{ZZ}(\s, \u) \Bigg \} ~, \label{fa+++0}
\eqa
\bqa
&& \delta^{vf}_{+-+0}(\beta_Z,\t,\u; m_f)/\Delta =
-\, \frac{4\s}{\s_4} (\t-\u -\beta_Z \s_4)C_0(\s)
-4 (\t-\u- \beta_Z \s) C_{ZZ}(\s)
\nonumber \\
&& + \frac{16 \beta_Z \mf^2 Y}{\s_4}\Fstu
-\, \frac{4 (1+\beta_Z)(\t+\mzd)}{\s_4}\Est
\nonumber \\
&& +\, \frac{4 (1-\beta_Z)(\u+\mzd)}{\s_4}\Esu ~,
 \label{fv+-+0}
\eqa
\bqa
&& \delta^{af}_{+-+0}(\beta_Z,\t,\u; m_f)/\Delta =
-\, \frac{4 \mf^2}{\mzd \s}(\t-\u +\beta_Z \s)
\Big \{ 2\s C_0(\s) +\Est +\Esu
\nonumber \\
&& +\frac{(\s +4 \mzd)Y}{\s_4} D_{ZZ}(\t,\u) \Big \}
+\frac{16 \beta_Z \mf^2 Y}{\s_4}\Fstu
-\frac{4 \s}{\s_4} (\t-\u -\beta_Z \s_4) C_0(\s)
\nonumber \\
&& - 4 (\t-\u - \beta_Z\s) C_{ZZ}(\s)
-\, \frac{4 (1+\beta_Z)(\t +\mzd)}{\s_4} \Est
\nonumber \\
&& + \,  \frac{4 (1-\beta_Z)(\u +\mzd)}{\s_4} \Esu
~ , \label{fa+-+0}
\eqa
\bqa
&& \delta^{vf}_{+-+-}(\beta_Z,\t,\u; m_f)=
- 4\s\Big [\frac{\s_2}{\s_4} C_0(\s) + C_{ZZ}(\s) \Big ]
- \frac{2}{\s_4} \Big [(\s_2+\beta_Z \s)\Est
\nonumber \\
&& + (\s_2-\beta_Z\s)\Esu \Big ]
-\, \frac{4\s\mf^2}{\s_4} [\s_4 +\beta_Z (\t-\u)]\Fstu
~ , \label{fv+-+-}
\eqa
\bqa
&& \delta^{af}_{+-+-}(\beta_Z,\t,\u; m_f)=
- 4\s\Big [\frac{\s_2}{\s_4} C_0(\s) + C_{ZZ}(\s) \Big ]
- \frac{2}{\s_4} \Big [(\s_2+\beta_Z \s)\Est
 + (\s_2-\beta_Z\s)\Esu \Big ]
\nonumber \\
&& -\, \frac{4\s\mf^2}{\s_4} [\s_4 +\beta_Z (\t-\u)]
[ \Fstu -2 D_{ZZ}(\t,\u)] - \frac{16 \mf^2
Y}{\s_4}D_{ZZ}(\t,\u) ~~ . \label{fa+-+-}
\eqa

We have checked that the fermion loop results in (\ref{LZff}-
\ref{fa+-+-}) agree with those of \cite{GloverZZ},
apart from the overall sign, provided that the replacement
\bq
\Delta_{ref.\cite{GloverZZ}}  ~ \longrightarrow
~ -~  \frac{2}{\s} ~ \Delta ~
\eq
is made\footnote{A factor of $\s$ is apparently missing
in the first term within the curly brackets in
Eqs.(3.14) of \cite{GloverZZ}.}.
In addition to this, it should be remembered
that our definitions of $\t$ and $\u$
should be interchanged when comparing with \cite{GloverZZ}, and
that these authors do not use the JW convention. \par


\newpage

\renewcommand{\theequation}{B.\arabic{equation}}
\renewcommand{\thesection}{B.\arabic{section}}
\setcounter{equation}{0}
\setcounter{section}{0}

{\large \bf Appendix B: The asymptotic
$\gamma \gamma \to Z Z $ amplitudes in SM.}

Since the expressions in Appendix A
for the $\gamma \gamma \to ZZ$ helicity
amplitudes  are rather complicated, it
would be useful to quote their asymptotic expressions
involving  logarithmic functions only. To this purpose, we need the
asymptotic expressions for the Passarino-Veltman functions in
(\ref{C0}, \ref{BZ}),
\bq
B_Z(\s) \simeq -\ln \left ( \frac{-\s- \ii \epsilon}
{-\mzd - \ii \epsilon}\right )  ~~ ,  \label{BZasym}
\eq
\bq
C_0(\s)  =  \frac{1}{2\s} \left [
\ln\left (\frac{\sqrt{ 1-~(4 m^2/\s) ~+\ii \epsilon}-1}
{\sqrt{ 1-~(4 m^2/\s) ~+\ii \epsilon}+1} \right )
\right ]^2 \simeq \frac{1}{2 \s}
\left [\ln \left (\frac{-\s -\ii\epsilon}{m^2}
\right ) \right ]^2 \ \ , \label{C0asym}
\eq
which should be valid for $|\s| \gg (m^2,~\mzd)$,
\cite{Denner, gggg, gggZ}.
\par

For the 1-loop functions
$C_Z(\s)$,  $C_{ZZ}$ and $D_{ZZ}$,
containing one or two legs at the
$Z$-mass shell, the asymptotic expressions
depend also
on the threshold singularity  through \cite{Denner}
\bq
a_Z \equiv   \sqrt{ 1-~\frac{4 m^2}{\mzd} ~+\ii \epsilon}
 ~ ~ . \label{aZ}
\eq
Taking then  $|\s| \gg (m^2,~\mzd)$, and using the  definition
\bq
\tilde a_Z \equiv
 \pi^2 - \left[ \ln \left(\frac{1+a_Z}{2}\right )
- \ln \left(\frac{1-a_Z}{2}\right ) \right ]^2
 +  2\ii\pi \left[ \ln \left(\frac{1+a_Z}{2}\right )
- \ln \left(\frac{1-a_Z}{2}\right )  \right ] ~ ,
\label{aZtilde}
\eq
we get
\bqa
C_Z(\s)& \simeq & \frac{1}{\s} \Bigg \{
\frac{1}{2}\ln^2\left (\frac{-\s-\ii\epsilon}{m^2}\right)
+ \frac{\tilde a_Z}{2}  \Bigg \} \ \ , \label{CZasym} \\
C_{ZZ}(\s)& \simeq & \frac{1}{\s} \Bigg \{
\frac{1}{2}\ln^2\left (\frac{-\s-\ii\epsilon}{m^2}\right)
+ \tilde a_Z  \Bigg \}  \ \ , \label{CZZasym}
\eqa
while for $(|\s|,~ |\u|) \gg (m^2,~\mzd)$, we get \cite{Denner}
\bqa
D_{ZZ}(\s, \t)& \simeq & \frac{2}{\s\t} \Bigg \{
\ln \left (\frac{-\s-\ii \epsilon}{m^2}\right)
\ln \left (\frac{-\t-\ii \epsilon}{m^2}\right)
- \, \frac{\pi^2}{2} + \tilde a_Z  \Bigg \} \ \  . \label{DZZasym}
\eqa \par

The principal value of the logarithms
is understood  in (\ref{BZasym}, \ref{C0asym},
\ref{CZasym}-\ref{DZZasym}), with the cuts  along the negative real axis.
These asymptotic expressions should be quite accurate
in the indicated regions, except in the case where $\mz \gg m$;
  which  leads  to   $a_Z \to 1$ and
$|\tilde a_Z| \to \infty $,  disturbing ( \ref{CZasym}-\ref{DZZasym}).
Thus, for  \eg\@  $C_{ZZ}(\s)$  at
 $\s=-\mz^4/m^2(1-4m^2/\mzd)\gg \mzd $,
the  exact expression in (\ref{CZZ}) differs considerably from
the asymptotic result of (\ref{CZZasym}).
Nevertheless,  it is shown below that these mass singularities
cancel in the asymptotic
behaviour of the physical $\gamma \gamma \to ZZ$ amplitudes.
A similar property has also been observed for the
$\gamma \gamma \to \gamma Z$ case \cite{gggZ}.
This cancellation should be a consequence of gauge invariance
and a reflection of the fact that although some single log
imaginary terms remain in the asymptotic expressions for the
physical amplitudes of these processes,
there are no overlapping soft and collinear singulaties which
 would had led to double-log Sudakov type terms  \cite{mass}.
We come back to this at the end of this Appendix.\par

Before turning to this though, we remark on the basis of
(\ref{CZasym}-\ref{DZZasym}), that for
$(\s \sim |\t| \sim |u|) \gg (\mzd, ~ m^2)$, the corresponding
asymptotic expressions for the functions in (\ref{Ftilde}
-\ref{E2}) are
\bqa
\tilde F(\s, \t,\u) & \simeq &
\frac{2}{\s\u} \ln \left (\frac{-\s-\ii\epsilon}{m^2}\right)
\ln \left (\frac{-\u-\ii\epsilon}{m^2}\right)
+ \frac{2}{\s\t} \ln \left (\frac{-\s-\ii\epsilon}{m^2}\right)
\ln \left (\frac{-\t-\ii\epsilon}{m^2}\right)
\nonumber \\
&&
+\frac{2}{\t\u} \ln \left (\frac{-\t-\ii\epsilon}{m^2}\right)
\ln \left (\frac{-\u-\ii\epsilon}{m^2}\right) \ ,
\label{Ftildeasym} \\
E_1(\s,\t) & \simeq & \pi^2 - \tilde a_Z  +
\ln^2\left ( \frac{-\t-\ii \epsilon}{m^2} \right )
-2 \ln\left ( \frac{-\s-\ii\epsilon}{m^2} \right )
\ln\left ( \frac{-\t-\ii\epsilon}{m^2} \right )
\ , \label{E1asym} \\
E_2(\t,\u) & \simeq & \pi^2  + \left [
\ln\left ( \frac{-\t-\ii\epsilon}{m^2} \right )
-\ln\left ( \frac{-\u-\ii\epsilon}{m^2} \right ) \right ]^2
\ . \label{E2asym}
\eqa \par

In the remaining part of this Appendix we give the asymptotic
expressions for the ten amplitudes in (\ref{8basic},
\ref{-beta++--}, \ref{-beta++-0}), by neglecting terms
of $O(\mzd/\s$  , $\mz m/\s$ ,  $ m^2/ \s) $. These should hold in
the region
\bq
\s \sim |\t| \sim |\u| \gg (\mzd ~, ~m^2) \label{asym-range}
\eq
where $m$ is the mass of the scalar or fermion particle
circulating in the loop. Thus, for the scalar loop
contributions, using (\ref{FSamp}) and
(\ref{S+++-}-\ref{S+-+-}), we find
\bqa
 A^S_{++++}& \simeq &
4 - \frac{4\u\t}{\s^2}
\left [ \ln^2 \Big | \frac{\t}{\u}\Big | +\pi^2 \right ]
+\frac{4(\t-\u)}{\s} \ln \Big | \frac{\t}{\u} \Big |
~ , \label{Sasym++++}\\
 A^S_{+-+-} & \simeq &
4 -\frac{4\s\t}{\u^2}
\left [ \ln^2 \Big | \frac{\s}{\t} \Big | - 2\ii \pi
 \ln \Big | \frac{\s}{\t} \Big | \right ]
+\frac{4(\s-\t)}{\u} \left [\ln \Big |\frac{\s}{\t} \Big | -
\ii \pi \right ] ~ , \label{Sasym+-+-} \\
 A^S_{+++0} &\simeq &
\sqrt{\frac{\s\mzd}{2\u\t}} \Bigg \{
-\frac{8(\t-\u)}{\s} + \frac{4(\t-\u)\t\u}{\s^3}
\left [ \ln^2 \Big | \frac{\t}{\u}\Big | +\pi^2 \right ]
+\frac{16\u\t}{\s^2} \ln \Big | \frac{\t}{\u}\Big |
\Bigg \} , \label{Sasym+++0} \\
A^S_{+-+0} &\simeq &
\sqrt{\frac{\s\mzd}{2\u\t}} \Bigg \{
-\frac{8\u}{\s}
+\frac{4\t}{\u}
\left [ \ln^2 \Big | \frac{\s}{\t} \Big | - 2\ii \pi
 \ln \Big | \frac{\s}{\t} \Big | \right ]
+\frac{8\t}{\s}
\left [\ln \Big |\frac{\s}{\t} \Big | - \ii \pi \right ]
\Bigg \} ~ , \label{Sasym+-+0}
\eqa
while the rest  are found to be very small, \ie
\bqa
A^S_{+++-} \simeq A^S_{++--} \simeq A^S_{+-++} & \simeq & -4 ~ ,
\label{Sasym1-small} \\
A^S_{+-00} \simeq  A^S_{++00}
\simeq A^S_{++-0}  & \simeq &   0   ~. \label{Sasym2-small}
\eqa\par

The corresponding asymptotic expressions for the $W$ loop
contributions are given by (\ref{FWamp}, \ref{deltaW}) and the
relations
\bqa
\delta^W_{++++}  &\simeq &
16 \cwd \Bigg \{
\ln^2 \Big | \frac{\t}{\u} \Big | +\pi^2
+\frac{\s}{\u}\ln \Big | \frac{\u}{\mwd} \Big |
\ln \Big | \frac{\s}{\t} \Big |
\nonumber \\
& + & \frac{\s}{\t}\ln \Big | \frac{\t}{\mwd} \Big |
\ln \Big | \frac{\s}{\u} \Big |
- \ii \pi \Big [
\frac{\s}{\u} \ln \Big | \frac{\u}{\mwd} \Big |
+ \frac{\s}{\t} \ln \Big | \frac{\t}{\mwd} \Big |\Big ]
\Bigg \} ~, \label{Wasym++++} \\
\delta^W_{+-+-}  &\simeq &
16 \cwd \Bigg \{
\ln^2 \Big | \frac{\t}{\s} \Big |
+\frac{\u}{\t}\ln \Big | \frac{\t}{\mwd} \Big |
\ln \Big | \frac{\u}{\s} \Big |
+\frac{\u}{\s}\ln \Big | \frac{\s}{\mwd} \Big |
\ln \Big | \frac{\u}{\t} \Big |
\nonumber \\
& + &  \ii \pi \Big [
\frac{\u}{\t} \ln \Big ( \frac{\s}{\mwd} \Big )
- \frac{\u}{\s} \ln \Big | \frac{\u}{\s} \Big |
-\frac{(\s^2+\t^2)}{\s \t} \ln \Big | \frac{\t}{\s} \Big |\Big ]
\Bigg \} ~, \label{Wasym+-+-} \\
\delta^W_{+-00} & \simeq &
-2 \Big [ \ln^2 \Big | \frac{\t}{\u} \Big |
+ \frac{\s}{\u}\ln^2 \Big | \frac{\t}{\s} \Big |
+\frac{\s}{\t} \ln^2 \Big | \frac{\u}{\s} \Big | \Big ]
-\frac{4}{\s} \ln \Big ( \frac{\s}{\mwd} \Big )
\Big [ \t \ln \Big | \frac{\t}{\s} \Big |
+ \u \ln \Big | \frac{\u}{\s} \Big | \Big ]
\nonumber \\
&+& \ii 4 \pi \Bigg \{
\ln \Big ( \frac{\s}{\mwd} \Big )
+\Big (1-\frac{\t^2}{\s\u} \Big ) \ln \Big | \frac{\t}{\s} \Big |
+\Big (1-\frac{\u^2}{\s\t} \Big ) \ln \Big | \frac{\u}{\s} \Big |
\Bigg \} ~, \label{Wasym+-00}\\
\delta^W_{++00}  & \simeq &
2 \ln^2 \Big ( \frac{\s}{\mwd} \Big ) -2 \pi^2
- \ii 4 \pi  \ln \Big ( \frac{\s}{\mwd} \Big ) ~ ,
\label{Wasym++00} \\
 \delta^W_{+++0} &\simeq &
\sqrt{\frac{\s\mzd}{2\u\t}} \Bigg \{
-\frac{2(\t-\u)}{\s} \Big [ \ln^2 \Big ( \frac{\s}{\mwd}\Big )
+ \ln^2 \Big | \frac{\t}{\u}\Big |
-\ii 2 \pi \ln \Big ( \frac{\s}{\mwd}\Big ) \Big ]
\nonumber \\
& - & 4 \Big [ \ln \Big ( \frac{\s}{\mwd}\Big ) - \ii \pi \Big ]
\ln \Big | \frac{\t}{\u}\Big | \Bigg \} ~, \label{Wasym+++0}\\
 \delta^W_{+-+0} &\simeq &
\sqrt{\frac{\s\mzd}{2\u\t}} \Bigg \{
\Big ( -\frac{2\u}{\s} +32 \cwd \frac{\t}{\s} \Big )
\Big [ \ln^2 \Big ( \frac{\s}{\mwd}\Big )
-\ii 2 \pi \ln \Big ( \frac{\s}{\mwd}\Big ) \Big ]
\nonumber \\
&+& 32 \cwd \Big [
\frac{\u}{\s}\ln \Big | \frac{\t}{\mwd}\Big |
\ln \Big | \frac{\u}{\mwd}\Big |
+\frac{\t}{\s} \ln^2 \Big | \frac{\t}{\mwd}\Big |
\Big ]
-\frac{2\u}{\s}\ln^2 \Big | \frac{\t}{\u}\Big |
\nonumber \\
&+& \Big [
-\frac{4\u(\t-\u)}{\s^2} +32 \cwd \frac{(\t^2+\s^2)}{\s^2} \Big ]
\Big [ \ln \Big ( \frac{\s}{\mwd}\Big ) -\ii \pi \Big ]
\ln \Big | \frac{\t}{\mwd}\Big |
\nonumber \\
&-& \frac{4 \u}{\s^2} (\u-\t -8 \cwd \t)
\Big [\ln \Big ( \frac{\s}{\mwd}\Big ) -\ii \pi \Big ]
\ln \Big | \frac{\u}{\mwd}\Big |
\Bigg \} ~ , \label{Wasym+-+0}
\eqa
\bq
\delta^W_{+++-} \simeq \delta^W_{++--}
\simeq \delta^W_{+-++} \simeq \delta^W_{++-0}
\simeq 0 ~ . \label{Wasym-small}
\eq

The asymptotic expressions for the fermion loop contributions may
be expressed from (\ref{Ffamp}, \ref{deltafv}, \ref{deltafa}) and
the results
\bqa
\delta^{vf}_{++++} \simeq &
\delta^{af}_{++++} \simeq & -4
\Bigg \{
\ln^2 \Big | \frac{\u}{\t}\Big | +\pi^2 \Bigg \} ~ ,
\label{fvasym++++} \\
\delta^{vf}_{+-+-} \simeq &
\delta^{af}_{+-+-} \simeq & -4
\Bigg \{
\ln^2 \Big | \frac{\s}{\t}\Big | -\ii 2\pi
\ln \Big | \frac{\s}{\t}\Big |  \Bigg \} ~ ,
\label{fvasym+-+-}
\eqa
\bqa
\delta^{vf}_{+-00} &\simeq & \delta^{vf}_{++00} \simeq
\delta^{vf}_{+++0} \simeq 0 ~ , \label{fvasym+-00} \\
\delta^{af}_{+-00} &\simeq &
-\, \frac{8 m^2}{\mzd} \Bigg \{
\frac{\t}{\u} \Big [\ln^2 \Big | \frac{\t}{\s}\Big |
+\ii 2\pi \ln \Big | \frac{\t}{\s}\Big | \Big ]
+ \frac{\u}{\t} \Big [\ln^2 \Big | \frac{\u}{\s}\Big |
+\ii 2\pi \ln \Big | \frac{\u}{\s}\Big | \Big ]
\Bigg \} ~ , \label{faasym+-00} \\
\delta^{af}_{++00} & \simeq &
-\, \frac{8 m^2}{\mzd} \Bigg \{
\ln^2 \Big ( \frac{\s}{m^2}\Big ) -\pi^2
-\ii 2\pi \ln \Big ( \frac{\s}{m^2}\Big )
\Bigg \} ~ , \label{faasym++00} \\
\delta^{af}_{+++0} & \simeq & -\, \sqrt{\frac{\s\mzd}{2\u\t}}
~ \frac{8 m^2}{\mzd} \Bigg \{  \frac{(\u-\t)}{\s}
\Big [
\ln^2 \Big ( \frac{\s}{m^2}\Big )
-\ii 2\pi \ln \Big ( \frac{\s}{m^2}\Big )
 + \ln^2  \Big | \frac{\t}{\u}\Big | \Big ]
\nonumber \\
&& - 2 \ln \Big ( \frac{\s}{m^2}\Big ) \ln \Big | \frac{\t}{\u}\Big |
+\ii 2 \pi \ln \Big | \frac{\t}{\u}\Big |
\Bigg \} ~ ,
\eqa
\bqa
\delta^{vf}_{+-+0} &\simeq &
-\, \sqrt{\frac{\s\mzd}{2\u\t}} ~ \frac{8 \t}{\s}
\Bigg \{
\ln^2 \Big | \frac{\s}{\t}\Big | -
\ii 2 \pi \ln \Big | \frac{\s}{\t}\Big | \Bigg \}
~ , \label{fvasym+-+0} \\
\delta^{af}_{+-+0} &\simeq &
-\, \sqrt{\frac{\s\mzd}{2\u\t}} ~ \frac{8 \t}{\s}
\Bigg \{ \ln^2 \Big | \frac{\s}{\t}\Big | -
\ii 2 \pi \ln \Big | \frac{\s}{\t}\Big | \Bigg \}
\nonumber \\
&& +\, \sqrt{\frac{\s \mzd }{2\u\t}} ~ \frac{8  m^2 \u}{\mzd \s}
\Bigg \{\ln^2 \Big ( \frac{\s m^2}{\t\u }\Big ) -
\ii 2 \pi \ln \Big ( \frac{\s m^2 }{\t\u}\Big ) \Bigg \}
~ , \label{faasym+-+0}
\eqa
\bq
\delta^{vf}_{+++-} \simeq \delta^{af}_{+++-}
\simeq \delta^{vf}_{++--} \simeq \delta^{af}_{++--}
\simeq \delta^{vf}_{+-++} \simeq \delta^{af}_{+-++}
\simeq \delta^{vf}_{++-0} \simeq \delta^{af}_{++-0}
\simeq 0 ~ . \label{fasym-small}
\eq

\vspace{1cm}

As promised, the asymptotic expressions
of the helicity amplitudes derived from the preceding formulae,
do not depend on the parameter $a_Z$ of
(\ref{aZ}, \ref{aZtilde}) entering
the Passarino-Veltman functions in
(\ref{CZasym}-\ref{DZZasym}, \ref{E1asym}).
We also notice that the Sudakov-type log-squared
terms in  (\ref{C0asym},  \ref{CZasym} - \ref{E2asym})
cancel out, when substituted to these  amplitudes, because of Bose
symmetry.
Therefore, in the asymptotic region indicated in
(\ref{asym-range}), the only large contributions
come from  the single-logarithm imaginary terms
appearing  in $\delta^W_{++++}$ and $\delta^W_{+-+-}$;
compare (\ref{Wasym++++} \ref{Wasym+-+-}).
These terms are the only ones which increase  (logarithmically)
with $\s$. \par

\clearpage
\newpage

\clearpage

\clearpage

\clearpage

\newpage
\clearpage

\clearpage

\addtocounter{figure}{-1}

\begin{figure}[p]
\vspace*{-4cm}
\[
\epsfig{file=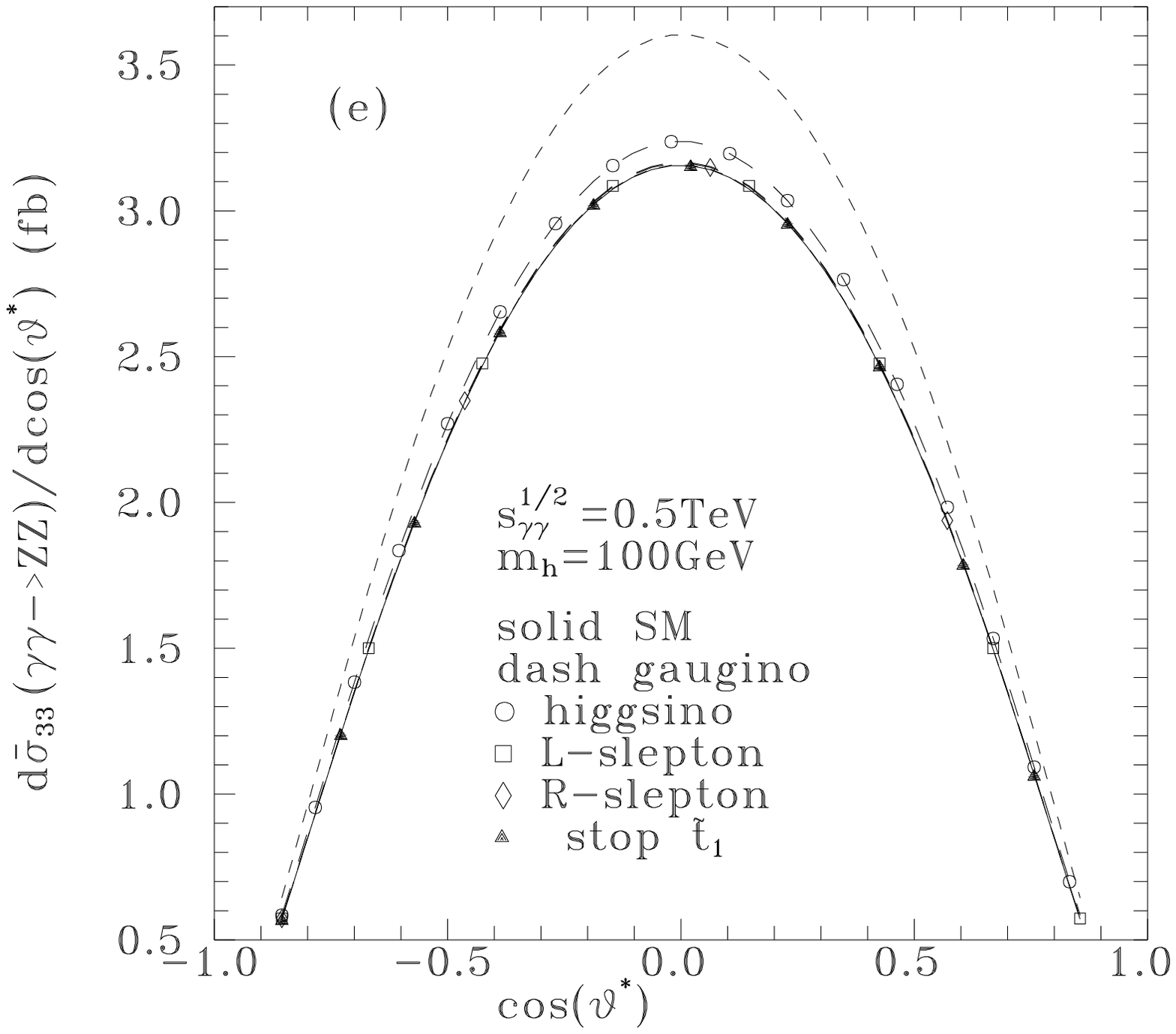,height=7.5cm}\hspace{0.5cm}
\epsfig{file=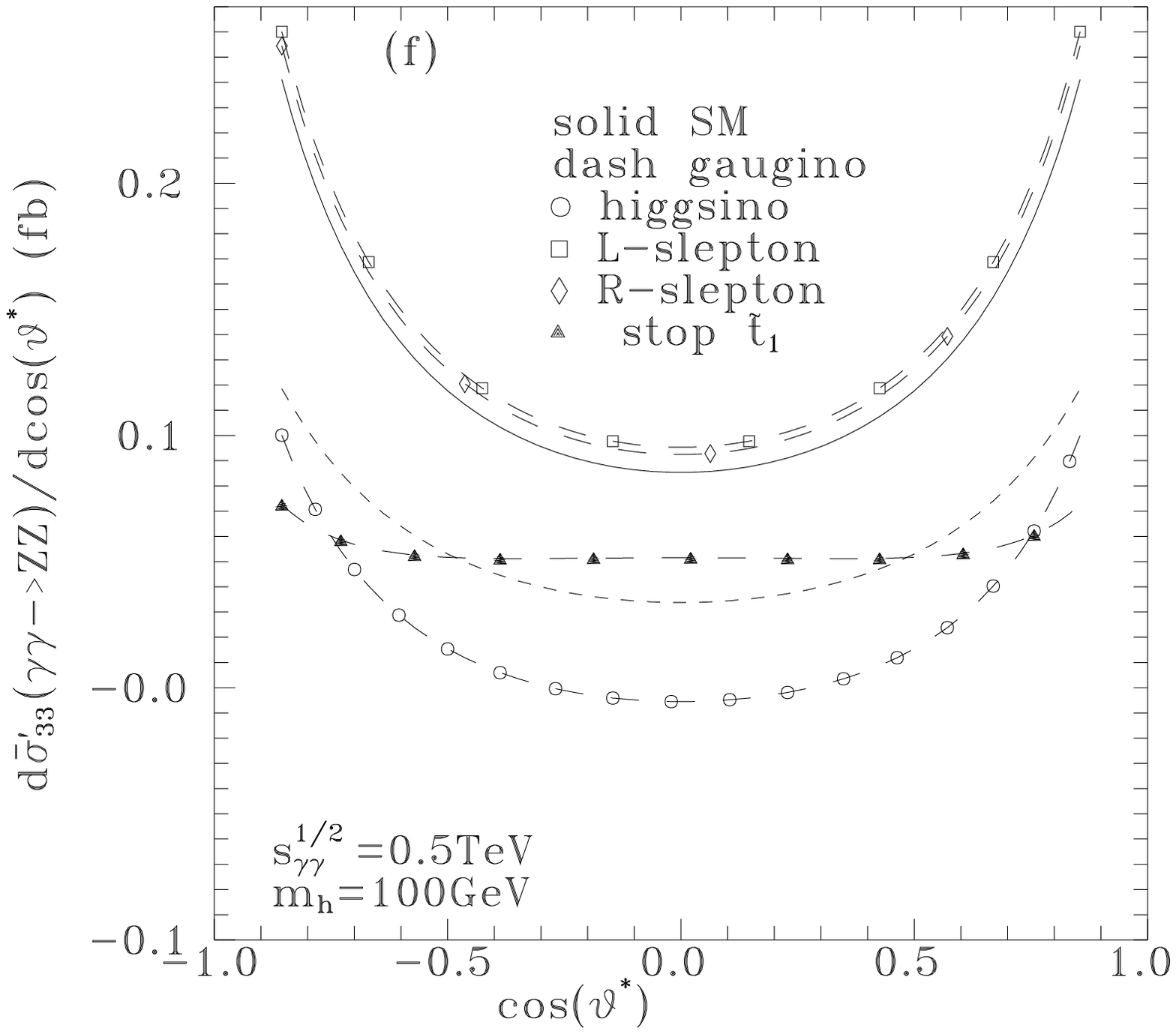,height=7.5cm}
\]
\vspace*{0.5cm}
\caption[1]{Angular distributions for
$d\bar{\sigma}_{33}/d\cos\vartheta^*$,
$d\bar{\sigma}^\prime_{33}/d\cos\vartheta^*$.}
\label{angular1}
\end{figure}

\clearpage

\begin{figure}[p]
\vspace*{-4cm}
\[
\epsfig{file=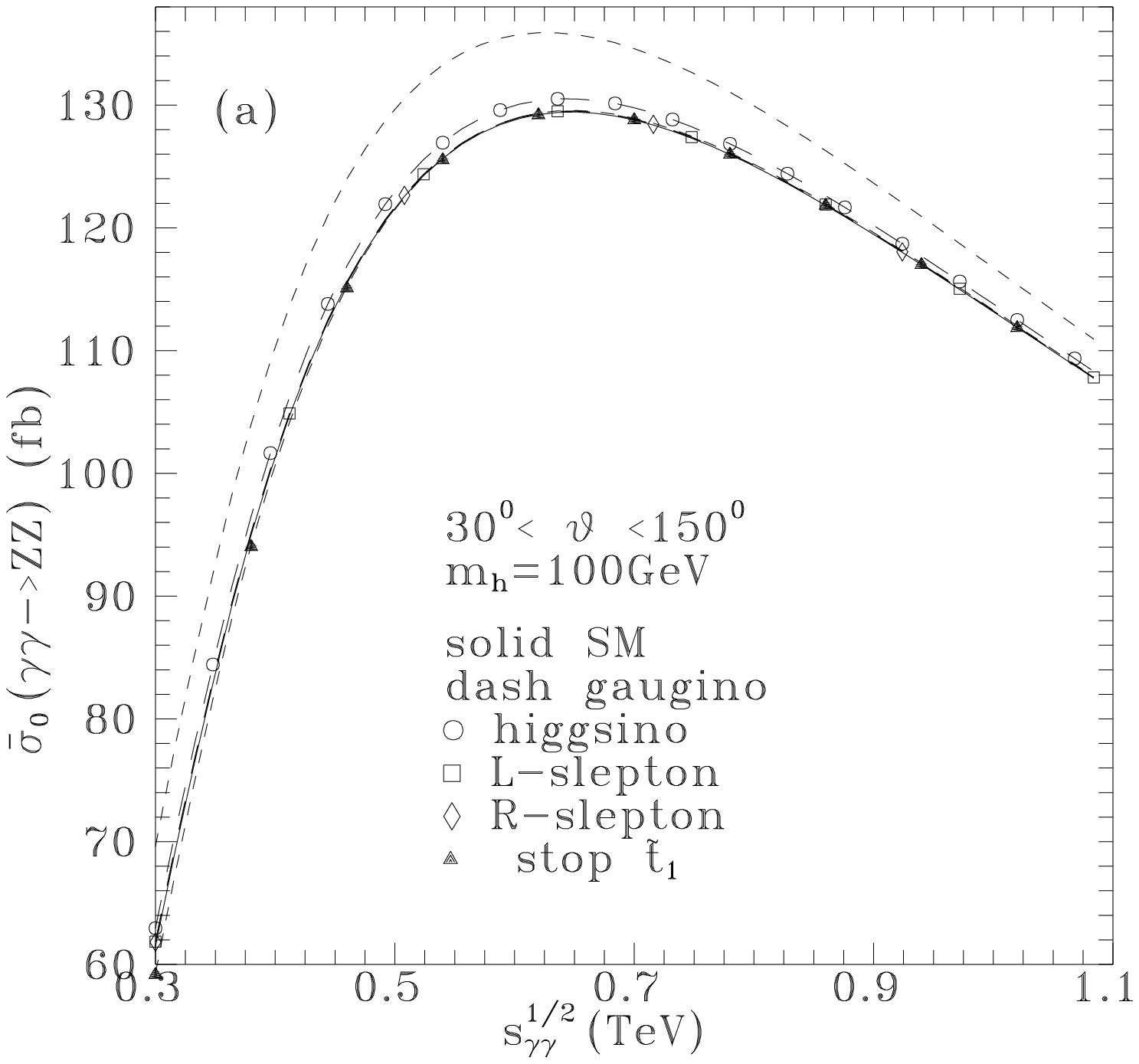,height=7.5cm}\hspace{0.5cm}
\epsfig{file=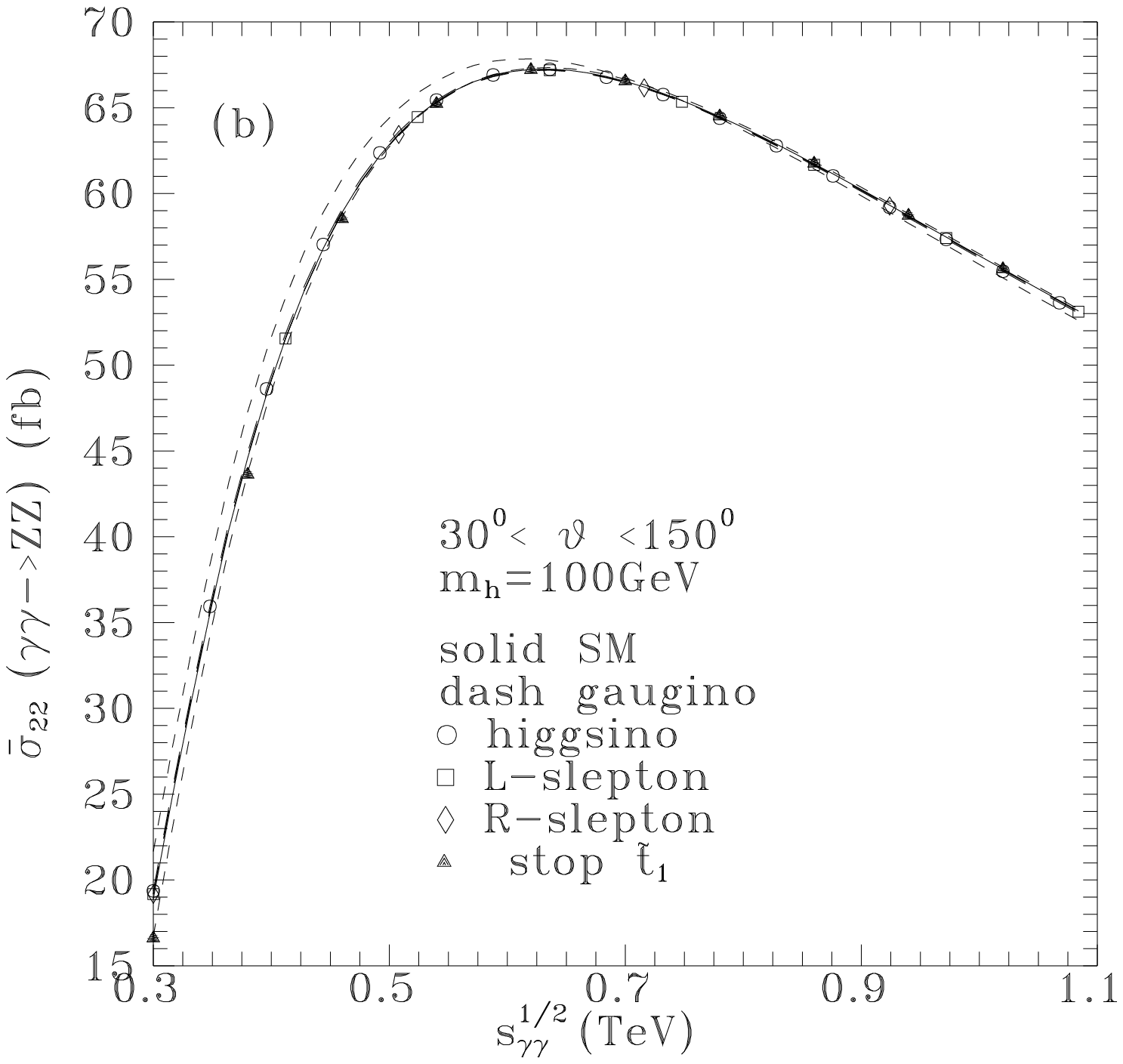,height=7.5cm}
\]
\vspace*{0.5cm}
\[
\epsfig{file=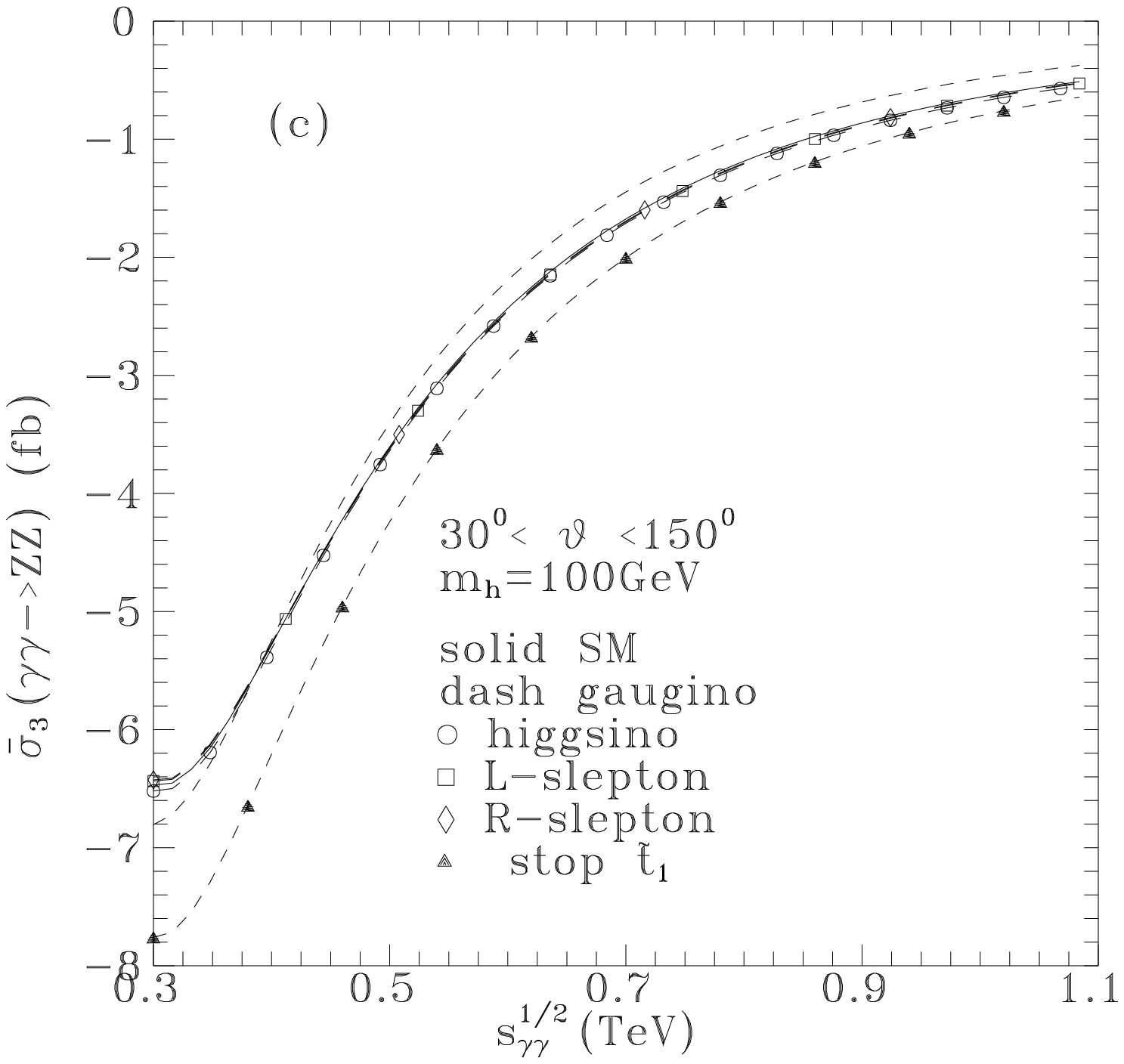,height=7.5cm}\hspace{0.5cm}
\epsfig{file=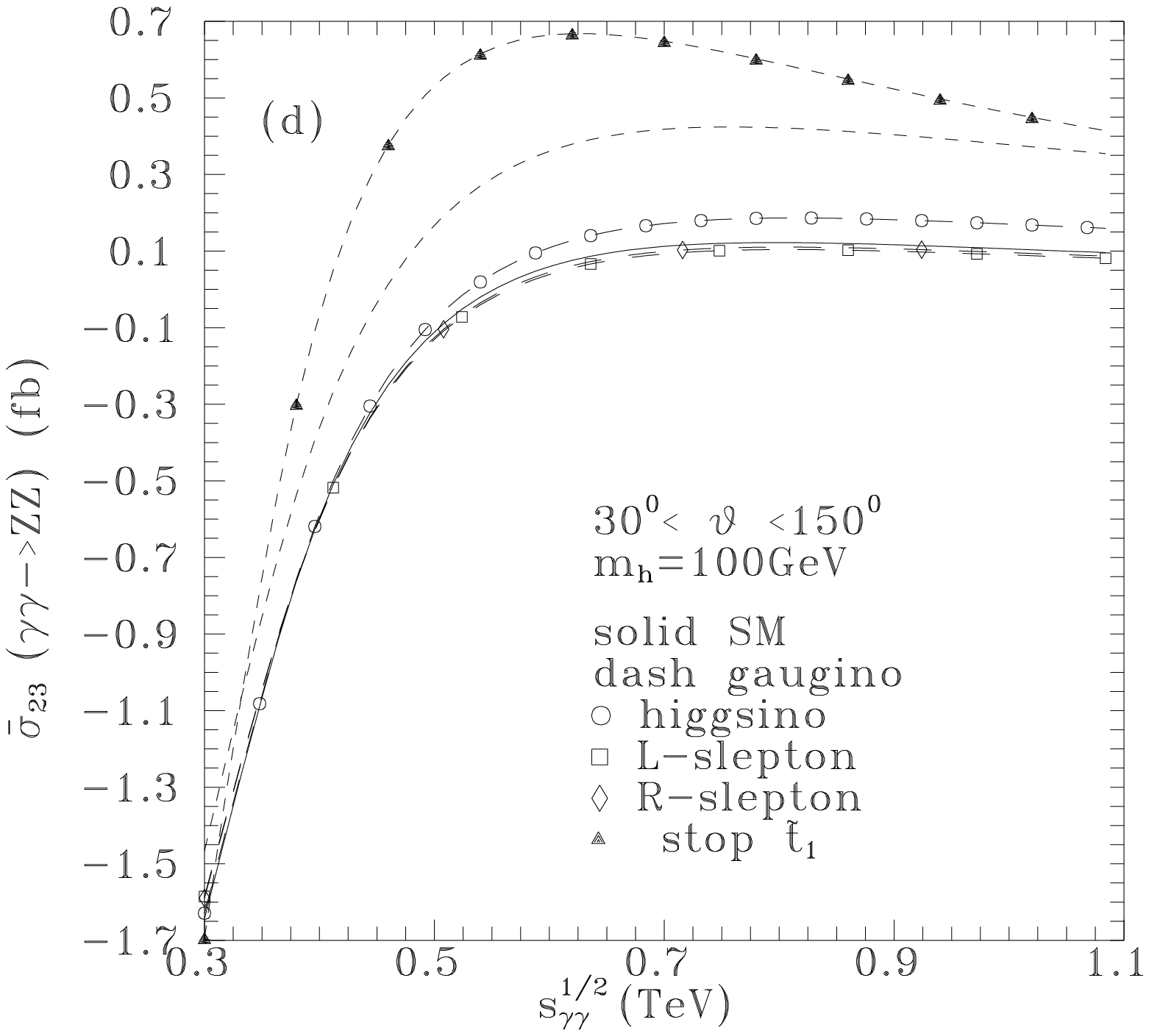,height=7.5cm}
\]
\vspace*{0.5cm}
\caption[1]{$\bar \sigma_0$, $\bar \sigma_{22}$,
$\bar \sigma_{3} $ and  $\bar \sigma_{23}$ for SM (solid) and
in the presence of a chargino,  or a selectron,
or a lightest stop contribution,
using the same parameters as in Fig.\ref{chargino-amp} or
Fig.\ref{slepton-amp}  or Fig.\ref{stop1-amp} respectively.}
\label{sig}
\end{figure}

\addtocounter{figure}{-1}

\begin{figure}[p]
\vspace*{-4cm}
\[
\epsfig{file=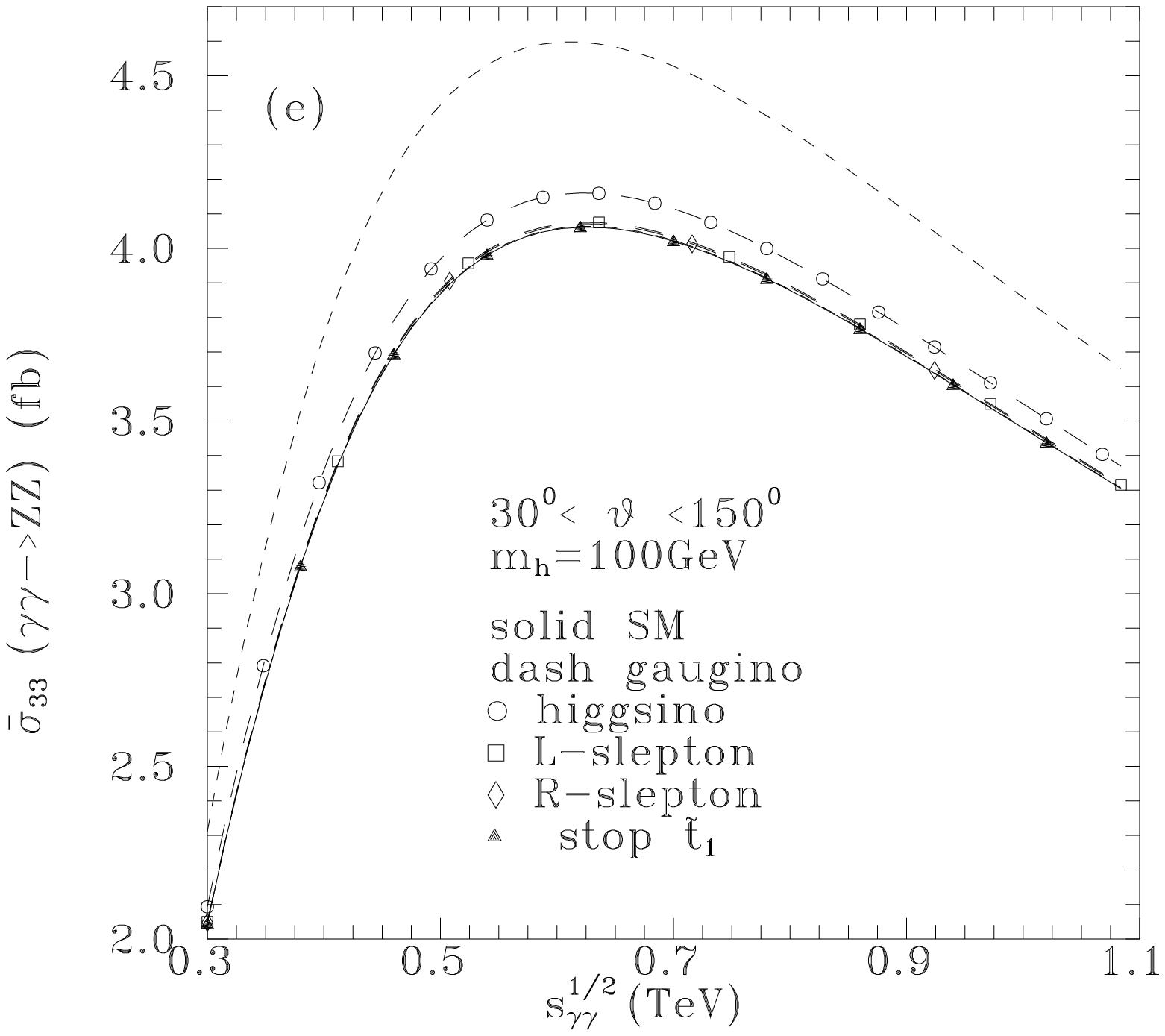,height=7.5cm}\hspace{0.5cm}
\epsfig{file=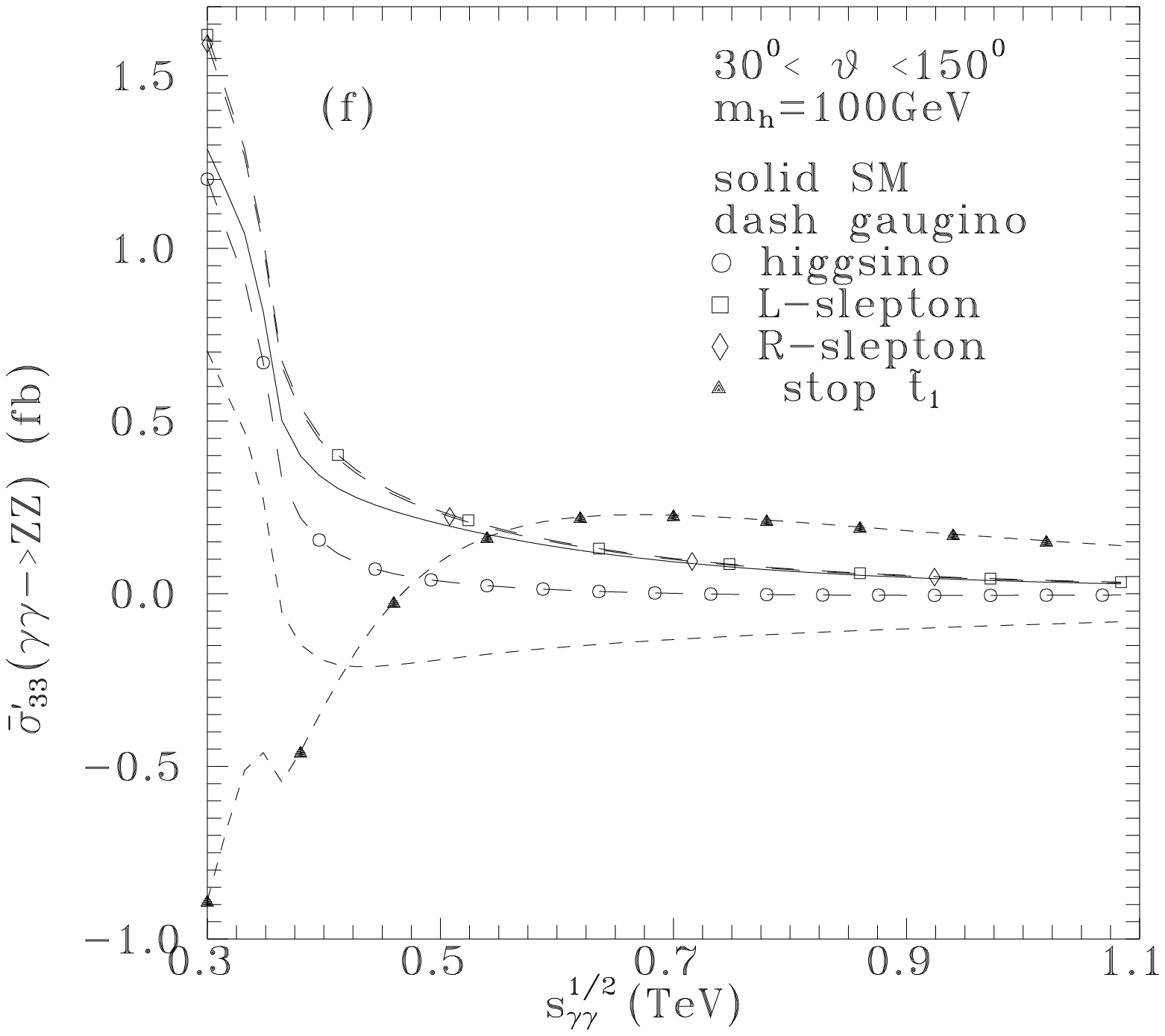,height=7.5cm}
\]
\caption[1]{$\bar \sigma_{33} $
and  $\bar \sigma'_{33} $ for SM (solid) and in the presence of
a chargino  or a selectron or a lightest stop contribution
using the same parameters as in
Fig.\ref{chargino-amp} or Fig.\ref{slepton-amp}  or
Fig.\ref{stop1-amp} respectively.}
\label{sig1}
\end{figure}

\end{document}